\documentclass[a4paper,twoside]{article}

\usepackage{graphicx,fancyhdr}

\usepackage{times}
 \rm

\title{The SSM Toolbox for Matlab}

\author{Jyh-Ying Peng$^{1,2}$, John A. D. Aston$^1$\\
\\$^1$Institute of Statistical Science,
     Academia Sinica\thanks{Address for Correspondence:\newline
Institute of Statistical Science,\newline
     Academia Sinica\newline
     128 Academia Road, Sec 2\newline
     Taipei 115, Taiwan\newline
          email: \{jypeng,jaston\}@stat.sinica.edu.tw}\\
$^2$ Computer Science and Information Engineering,
    National Taiwan University
}

\begin{document}

\maketitle
\thispagestyle{empty}

\newpage

\tableofcontents

\newpage

\begin{abstract}
State Space Models (SSM) is a MATLAB 7.0 software toolbox for doing time series analysis by state space methods. The software features fully interactive construction and combination of models, with support for univariate and multivariate models, complex time-varying (dynamic) models, non-Gaussian models, and various standard models such as ARIMA and structural time-series models. The software includes standard functions for Kalman filtering and smoothing, simulation smoothing, likelihood evaluation, parameter estimation, signal extraction and forecasting, with incorporation of exact initialization for filters and smoothers, and support for missing observations and multiple time series input with common analysis structure. The software also includes implementations of TRAMO model selection and Hillmer-Tiao decomposition for ARIMA models. The software will provide a general toolbox for doing time series analysis on the MATLAB platform, allowing users to take advantage of its readily available graph plotting and general matrix computation capabilities.

\vskip6pt

\end{abstract}

\newpage

\section{Introduction}

State Space Models (SSM) is a MATLAB toolbox for doing time series analysis using general state space models and the Kalman filter \cite{Durbin01}. The goal of this software package is to provide users with an intuitive, convenient and efficient way to do general time series modeling within the state space framework. Specifically, it seeks to provide users with easy construction and combination of arbitrary models without having to explicitly define every component of the model, and to maximize transparency in their data analysis usage so no special consideration is needed for any individual model. This is achieved through the unification of all state space models and their extension to non-Gaussian and nonlinear special cases. The user creation of custom models is also implemented to be as general, flexible and efficient as possible. Thus, there are often multiple ways of defining a single model and choices as to the  parametrization versus initialization and to how the model update functions are implemented. Stock model components are also provided to ease user extension to existing predefined models. Functions that implement standard algorithms such as the Kalman filter and state smoother, log likelihood calculation and parameter estimation will work seamlessly across all models, including any user defined custom models.

These features are implemented through object-oriented programming primitives provided by MATLAB and classes in the toolbox are defined to conform to MATLAB conventions whenever possible. The result is a highly integrated toolbox with support for general state space models and standard state space algorithms, complemented by the built-in matrix computation and graphic plotting capabilities of MATLAB.

\section{The State Space Models}

This section presents a summary of the basic definition of models supported by SSM. Currently SSM implements the Kalman filter and related algorithms for model and state estimation, hence non-Gaussian or nonlinear models need to be approximated by linear Gaussian models prior to or during estimation. However the approximation is done automatically and seamlessly by the respective routines, even for user-defined non-Gaussian or nonlinear models.

The following notation for various sequences will be used throughout the paper:\medskip\\
\begin{tabular}{ccl}
$y_t$           & $p\times1$ & Observation sequence \\
$\varepsilon_t$ & $p\times1$ & Observation disturbance (Unobserved) \\
$\alpha_t$      & $m\times1$ & State sequence (Unobserved) \\
$\eta_t$        & $r\times1$ & State disturbance (Unobserved)
\end{tabular}

\subsection{Linear Gaussian models}

SSM supports linear Gaussian state space model in the form
\begin{equation}
\begin{array}{rcll}
y_t          & =    & Z_t\alpha_t + \varepsilon_t,   & \varepsilon_t \sim \mathcal{N}(0, H_t), \\
\alpha_{t+1} & =    & c_t + T_t\alpha_t + R_t\eta_t, & \eta_t        \sim \mathcal{N}(0, Q_t), \\
\alpha_1     & \sim & \mathcal{N}(a_1, P_1),         & t             =    1,\ldots,n.
\end{array}
\end{equation}
Thus the matrices $Z_t, c_t, T_t, R_t, H_t, Q_t, a_1, P_1$ are required to define a linear Gaussian state space model. The matrix $Z_t$ is the state to observation linear transformation, for univariate models it is a row vector. The matrix $c_t$ is the same size as the state vector, and is the ``constant'' in the state update equation, although it can be dynamic or dependent on model parameters. The square matrix $T_t$ defines the time evolution of states. The matrix $R_t$ transforms general disturbance into state space, and exists to allow for more varieties of models. $H_t$ and $Q_t$ are Gaussian variance matrices governing the disturbances, and $a_1$ and $P_1$ are the initial conditions. The matrices and their dimensions are listed:\medskip\\
\begin{tabular}{ccl}
$Z_t$ & $p\times m$ & State to observation transform matrix \\
$c_t$ & $m\times1$  & State update constant \\
$T_t$ & $m\times m$ & State update transform matrix \\
$R_t$ & $m\times r$ & State disturbance transform matrix \\
$H_t$ & $p\times p$ & Observation disturbance variance \\
$Q_t$ & $r\times r$ & State disturbance variance \\
$a_1$ & $m\times1$  & Initial state mean \\
$P_1$ & $m\times m$ & Initial state variance
\end{tabular}

\subsection{Non-Gaussian models}

SSM supports non-Gaussian state space models in the form
\begin{equation}
\begin{array}{rcll}
y_t          & \sim & p(y_t|\theta_t),               & \theta_t=Z_t\alpha_t, \\
\alpha_{t+1} & =    & c_t + T_t\alpha_t + R_t\eta_t, & \eta_t\sim Q_t=p(\eta_t), \\
\alpha_1     & \sim & \mathcal{N}(a_1, P_1),         & t=1,\ldots,n.
\end{array}
\end{equation}
The sequence $\theta_t$ is the signal and $Q_t$ is a non-Gaussian distribution (e.g. heavy-tailed distribution). The non-Gaussian observation disturbance can take two forms: an exponential family distribution
\begin{equation}
H_t = p(y_t|\theta_t) = \exp\left[y_t^T\theta_t-b_t(\theta_t)+c_t(y_t)\right], \; -\infty<\theta_t<\infty,
\end{equation}
or a non-Gaussian additive noise
\begin{equation}
y_t = \theta_t + \varepsilon_t, \; \varepsilon_t \sim H_t = p(\varepsilon_t).
\end{equation}
With model combination it is also possible for $H_t$ and $Q_t$ to be a combination of Gaussian distributions (represented by variance matrices) and various non-Gaussian distributions.

\subsection{Nonlinear models}

SSM supports nonlinear state space models in the form
\begin{equation}
\begin{array}{rcll}
y_t          & =    & Z_t(\alpha_t) + \varepsilon_t,   & \varepsilon_t\sim\mathcal{N}(0, H_t), \\
\alpha_{t+1} & =    & c_t + T_t(\alpha_t) + R_t\eta_t, & \eta_t\sim\mathcal{N}(0, Q_t), \\
\alpha_1     & \sim & \mathcal{N}(a_1, P_1),           & t=1,\ldots,n.
\end{array}
\end{equation}
$Z_t$ and $T_t$ are functions that map $m\times1$ vectors to $p\times1$ and $m\times1$ vectors respectively. With model combination it is also possible for $Z_t$ and $T_t$ to be a combination of linear functions (matrices) and nonlinear functions.

\section{The State Space Algorithms}

This section summarizes the core algorithms implemented in SSM, starting with the Kalman filter. Normally the initial conditions $a_1$ and $P_1$ should be given for these algorithms, but the exact diffuse initialization implemented permits elements of either to be unknown, and derived implicitly from the first few observations. The unknown elements in $a_1$ are set to $0$, and the corresponding diagonal entry in $P_1$ is set to $\infty$, this represents an improper prior which reflects the lack of knowledge about that particular element \emph{a priori}. See \cite{Durbin01} for details. To keep variables finite in the algorithms we define another initial condition $P_{\infty,1}$ with elements
\begin{equation}
\left(P_{\infty,1}\right)_{i,j} = \left\{\begin{array}{cl}0, & \mbox{if } \left(P_1\right)_{i,j}<\infty, \\
                                                          1, & \mbox{if } \left(P_1\right)_{i,j}=\infty, \end{array}\right.
\end{equation}
and set infinite elements of $P_1$ to $0$. All the state space algorithms supports exact diffuse initialization.

For non-Gaussian and nonlinear models linear Gaussian approximations are performed by linearization of the loglikelihood equation and model matrices respectively, and the Kalman filter algorithms can then be used on the resulting approximation models.

\subsection{Kalman filter}

Given the observation $y_t$, the model and initial conditions $a_1, P_1, P_{\infty,1}$, Kalman filter outputs $a_t=\mathrm{E}(\alpha_t|y_{1:t-1})$ and $P_t=\mathrm{Var}(\alpha_t|y_{1:t-1})$, the one step ahead state prediction and prediction variance. Theoretically the diffuse state variance $P_{\infty,t}$ will only be non-zero for the first couple of time points (usually the number of diffuse elements, but may be longer depending on the structure of $Z_t$), after which exact diffuse initialization is completed and normal Kalman filter is applied. Here $n$ denotes the length of observation data, and $d$ denotes the number of time points where the state variance remains diffuse.

The normal Kalman filter recursion (for $t>d$) are as follows:
\begin{eqnarray*}
  v_t &=& y_t-Z_ta_t, \\
  F_t &=& Z_tP_tZ_t^T+H_t, \\
  K_t &=& T_tP_tZ_t^TF_t^{-1}, \\
  L_t &=& T_t-K_tZ_t,  \\
  a_{t+1} &=& c_t+T_ta_t+K_tv_t, \\
  P_{t+1} &=& T_tP_tL_t^T + R_tQ_tR_t^T.
\end{eqnarray*}
The exact diffuse initialized Kalman filter recursion (for $t\leq d$) is divided into two cases, when $F_{\infty,t}=Z_tP_{\infty,t}Z_t^T$ is nonsingular:
\begin{eqnarray*}
  v_t       &=& y_t-Z_ta_t, \\
  F_t^{(1)} &=& (Z_tP_{\infty,t}Z_t^T)^{-1}, \\
  F_t^{(2)} &=& -F_t^{(1)}(Z_tP_tZ_t^T+H_t)F_t^{(1)}, \\
  K_t       &=& T_tP_{\infty,t}Z_t^TF_t^{(1)}, \\
  K_t^{(1)} &=& T_t(P_tZ_t^TF_t^{(1)}+P_{\infty,t}Z_t^TF_t^{(2)}), \\
  L_t       &=& T_t-K_tZ_t, \\
  L_t^{(1)} &=& -K_t^{(1)}Z_t, \\
  a_{t+1}        &=& c_t+T_ta_t+K_tv_t, \\
  P_{t+1}        &=& T_tP_{\infty,t}L_t^{(1)T}+T_tP_tL_t^T+R_tQ_tR_t^T, \\
  P_{\infty,t+1} &=& T_tP_{\infty,t}L_t^T,
\end{eqnarray*}
and when $F_{\infty,t}=0$:
\begin{eqnarray*}
  v_t &=& y_t-Z_ta_t, \\
  F_t &=& Z_tP_tZ_t^T+H_t, \\
  K_t &=& T_tP_tZ_t^TF_t^{-1}, \\
  L_t &=& T_t-K_tZ_t, \\
  a_{t+1}        &=& c_t+T_ta_t+K_tv_t, \\
  P_{t+1}        &=& T_tP_tL_t^T+R_tQ_tR_t^T, \\
  P_{\infty,t+1} &=& T_tP_{\infty,t}T_t^T.
\end{eqnarray*}

\subsection{State smoother}

Given the Kalman filter outputs, the state smoother outputs $\hat\alpha_t=\mathrm{E}(\alpha_t|y_{1:n})$ and $V_t=\mathrm{Var}(\alpha_t|y_{1:n})$, the smoothed state mean and variance given the whole observation data, by backwards recursion.

To start the backwards recursion, first set $r_n=0$ and $N_n=0$, where $r_t$ and $N_t$ are the same size as $a_t$ and $P_t$ respectively. The normal backwards recursion (for $t>d$) is then
\begin{eqnarray*}
  r_{t-1}        &=& Z_t^TF_t^{-1}v_t+L_t^Tr_t, \\
  N_{t-1}        &=& Z_t^TF_t^{-1}Z_t+L_t^TN_tL_t, \\
  \hat{\alpha}_t &=& a_t+P_tr_{t-1}, \\
  V_t            &=& P_t - P_tN_{t-1}P_t.
\end{eqnarray*}
For the exact diffuse initialized portion ($t\leq d$) set additional diffuse variables $r_d^{(1)}=0$ and $N_d^{(1)}=N_d^{(2)}=0$ corresponding to $r_d$ and $N_d$ respectively, the backwards recursion is
\begin{eqnarray*}
  r_{t-1}^{(1)} &=& Z_t^TF_t^{(1)}v_t+L_t^Tr_t^{(1)}+L_t^{(1)T}r_t, \\
  r_{t-1}       &=& L_t^Tr_t, \\
  N_{t-1}^{(2)} &=& Z_t^TF_t^{(2)}Z_t+L_t^TN_t^{(2)}L_t+L_t^TN_t^{(1)}L_t^{(1)} \\
                & & {}+L_t^{(1)T}N_t^{(1)}L_t+L_t^{(1)T}N_tL_t^{(1)}, \\
  N_{t-1}^{(1)} &=& Z_t^TF_t^{(1)}Z_t+L_t^TN_t^{(1)}L_t+L_t^{(1)T}N_tL_t, \\
  N_{t-1}       &=& L_t^TN_tL_t, \\
  \hat{\alpha}_t &=& a_t+P_tr_{t-1}+P_{\infty,t}r_{t-1}^{(1)}, \\
  V_t            &=& P_t-P_tN_{t-1}P_t-(P_{\infty,t}N_{t-1}^{(1)}P_t)^T \\
                 & & {}-P_{\infty,t}N_{t-1}^{(1)}P_t-P_{\infty,t}N_{t-1}^{(2)}P_{\infty,t},
\end{eqnarray*}
if $F_{\infty,t}$ is nonsingular, and
\begin{eqnarray*}
  r_{t-1}^{(1)} &=& T_t^Tr_t^{(1)}, \\
  r_{t-1}       &=& Z_t^TF_t^{-1}v_t+L_t^Tr_t, \\
  N_{t-1}^{(2)} &=& T_t^TN_t^{(2)}T_t, \\
  N_{t-1}^{(1)} &=& T_t^TN_t^{(1)}L_t, \\
  N_{t-1}       &=& Z_t^TF_t^{-1}Z_t+L_t^TN_tL_t, \\
  \hat{\alpha}_t &=& a_t+P_tr_{t-1}+P_{\infty,t}r_{t-1}^{(1)}, \\
  V_t            &=& P_t-P_tN_{t-1}P_t-(P_{\infty,t}N_{t-1}^{(1)}P_t)^T \\
                 & & {}-P_{\infty,t}N_{t-1}^{(1)}P_t-P_{\infty,t}N_{t-1}^{(2)}P_{\infty,t},
\end{eqnarray*}
if $F_{\infty,t}=0$.

\subsection{Disturbance smoother}

Given the Kalman filter outputs, the disturbance smoother outputs $\hat\varepsilon_t=\mathrm{E}(\varepsilon_t|y_{1:n})$, $\mathrm{Var}(\varepsilon_t|y_{1:n})$, $\hat\eta_t=\mathrm{E}(\eta_t|y_{1:n})$ and $\mathrm{Var}(\eta_t|y_{1:n})$. Calculate the sequence $r_t$ and $N_t$ by backwards recursion as before, the disturbance can then be estimated by
\begin{eqnarray*}
  \hat{\eta}_t                 &=& Q_tR_t^Tr_t, \\
  \mathrm{Var}(\eta_t|y_{1:n}) &=& Q_t-Q_tR_t^TN_tR_tQ_t, \\
  \hat{\varepsilon}_t                 &=& H_t(F_t^{-1}v_t-K_t^Tr_t), \\
  \mathrm{Var}(\varepsilon_t|y_{1:n}) &=& H_t-H_t(F_t^{-1}+K_t^TN_tK_t)H_t,
\end{eqnarray*}
if $t>d$ or $F_{\infty,t}=0$, and
\begin{eqnarray*}
  \hat{\eta}_t                 &=& Q_tR_t^Tr_t, \\
  \mathrm{Var}(\eta_t|y_{1:n}) &=& Q_t-Q_tR_t^TN_tR_tQ_t, \\
  \hat{\varepsilon}_t                 &=& -H_tK_t^Tr_t, \\
  \mathrm{Var}(\varepsilon_t|y_{1:n}) &=& H_t-H_tK_t^TN_tK_tH_t,
\end{eqnarray*}
if $t\leq d$ and $F_{\infty,t}$ is nonsingular.

\subsection{Simulation smoother}

The simulation smoother randomly generates an observation sequence $\tilde{y}_t$ and the underlying state and disturbance sequences $\tilde{\alpha}_t$, $\tilde{\epsilon}_t$ and $\tilde{\eta}_t$ conditional on the model and observation data $y_t$. The algorithm is as follows (each step is performed for all time points $t$):
\begin{enumerate}
  \item Draw random samples of the state and observation disturbances:
        \begin{eqnarray*}
        \varepsilon_t^+ & \sim & \mathcal{N}(0,H_t), \\
        \eta_t^+        & \sim & \mathcal{N}(0,Q_t).
        \end{eqnarray*}
  \item Draw random samples of the initial state:
        \[\alpha_1^+ \sim \mathcal{N}(a_1, P_1).\]
  \item Generate state sequences $\alpha_t^+$ and observation sequences $y_t^+$ from the random samples.
  \item Use disturbance smoothing to obtain
        \begin{eqnarray*}
        \hat{\varepsilon}_t   & = & \mathrm{E}(\varepsilon_t|y_{1:n}), \\
        \hat{\eta}_t          & = & \mathrm{E}(\eta_t|y_{1:n}), \\
        \hat{\varepsilon}_t^+ & = & \mathrm{E}(\varepsilon_t^+|y_{1:n}^+), \\
        \hat{\eta}_t^+        & = & \mathrm{E}(\eta_t^+|y_{1:n}^+).
        \end{eqnarray*}
  \item Use state smoothing to obtain
        \begin{eqnarray*}
        \hat{\alpha}_t   & = & \mathrm{E}(\alpha_t|y_{1:n}), \\
        \hat{\alpha}_t^+ & = & \mathrm{E}(\alpha_t^+|y_{1:n}^+).
        \end{eqnarray*}
  \item Calculate
        \begin{eqnarray*}
        \tilde{\alpha}_t      & = & \hat{\alpha}_t - \hat{\alpha}_t^+ + \alpha_t^+, \\
        \tilde{\varepsilon}_t & = & \hat{\varepsilon}_t - \hat{\varepsilon}_t^+ + \varepsilon_t^+, \\
        \tilde{\eta}_t        & = & \hat{\eta}_t - \hat{\eta}_t^+ + \eta_t^+.
        \end{eqnarray*}
  \item Generate $\tilde{y}_t$ from the output.
\end{enumerate}

\section{Getting Started}

The easiest and most frequent way to start using SSM is by constructing predefined models, as opposed to creating a model from scratch. This section presents some examples of simple time series analysis using predefined models, the complete list of available predefined models can be found in appendix \ref{app:predefined}.

\subsection{Preliminaries}

Some parts of SSM is programmed in c for maximum efficiency, before using SSM, go to the subfolder \texttt{src} and execute the script \texttt{mexall} to compile and distribute all needed c mex functions.

\subsection{Model construction}

To create an instance of a predefined model, call the SSMODEL (state space model) constructor with a coded string representing the model as the first argument, and optional arguments as necessary. For example:
\begin{itemize}
\item \texttt{model = ssmodel('llm')} creates a local level model.
\item \texttt{model = ssmodel('arma', p, q)} creates an ARMA$(p,q)$ model.
\end{itemize}
The resulting variable \texttt{model} is a SSMODEL object and can be displayed just like any other MATLAB variables. To set or change the model parameters, use \texttt{model.param}, which is a row vector that behaves like a MATLAB matrix except its size cannot be changed. The initial conditions usually defaults to exact diffuse initialization, where \texttt{model.a1} is zero, and \texttt{model.P1} is $\infty$ on the diagonals, but can likewise be changed. Models can be combined by horizontal concatenation, where only the observation disturbance model of the first one will be retained. The class SSMODEL will be presented in detail in later sections.

\subsection{Model and state estimation}

With the model created, estimation can be performed. SSM expects the data \texttt{y} to be a matrix of dimensions $p\times n$, where $n$ is the data size (or time duration). The model parameters are estimated by maximum likelihood, the SSMODEL class method \texttt{estimate} performs the estimation. For example:
\begin{itemize}
\item \texttt{model1 = estimate(y, model0)} estimates the model and stores the result in \texttt{model1}, where the parameter values of \texttt{model0} is used as initial value.
\item \texttt{[model1 logL] = estimate(y, model0, psi0, [], optname1,\\optvalue1, optname2, optvalue2, ...)} estimates the model with \texttt{psi0} as the initial parameters using option settings specified with option value pairs, and returns the resulting model and loglikelihood.
\end{itemize}

After the model is estimated, state estimation can be performed, this is done by the SSMODEL class method \texttt{kalman} and \texttt{statesmo}, which is the Kalman filter and state smoother respectively.
\begin{itemize}
\item \texttt{[a P] = kalman(y, model)} applies the Kalman filter on \texttt{y} and returns the one-step-ahead state prediction and variance.
\item \texttt{[alphahat V] = statesmo(y, model)} applies the state smoother on \texttt{y} and returns the expected state mean and variance.
\end{itemize}
The filtered and smoothed state sequences \texttt{a} and \texttt{alphahat} are \texttt{m}${}\times{}$\texttt{n+1} and \texttt{m}${}\times{}$\texttt{n} matrices respectively, and the filtered and smoothed state variance sequences \texttt{P} and \texttt{V} are \texttt{m}${}\times{}$\texttt{m}${}\times{}$\texttt{n+1} and \texttt{m}${}\times{}$\texttt{m}${}\times{}$\texttt{n} matrices respectively, except if \texttt{m}${}=1$, in which case they are squeezed and transposed.

\section{SSM Classes}

SSM is composed of various MATLAB classes each of which represents specific parts of a general state space model. The class \texttt{SSMODEL} represents a state space model, and is the most important class of SSM. It embeds the classes \texttt{SSMAT}, \texttt{SSDIST}, \texttt{SSFUNC} and \texttt{SSPARAM}, all designed to facilitate the construction and usage of \texttt{SSMODEL} for general data analysis. Consult appendix \ref{app:class} for complete details about these classes.

\subsection{Class SSMAT}

The class state space matrix (\texttt{SSMAT}) forms the basic components of a state space model, they can represent linear transformations that map vectors to vectors (the $Z_t$, $T_t$ and $R_t$ matrices), or variance matrices of Gaussian disturbances (the $H_t$ and $Q_t$ matrices).

The first data member of class \texttt{SSMAT} is \texttt{mat}, which is a matrix that represents the stationary part of the state space matrix, and is often the only data member that is needed to represent a state space matrix. For dynamic state space matrices, the additional data members \texttt{dmmask} and \texttt{dvec} are needed. \texttt{dmmask} is a logical matrix the same size as \texttt{mat}, and specifies elements of the state space matrix that varies with time, these elements arranged in a column in column-wise order forms the columns of the matrix \texttt{dvec}, thus \texttt{nnz(dmmask)} is equal to the number of rows in \texttt{dvec}, and the number of columns of \texttt{dvec} is the number of time points specified for the state space matrix. This is designed so that the code
\begin{center}\tt mat(dmmask) = dvec(:, t);\end{center}
gets the value of the state space matrix at time point \texttt{t}.

Now these are the constant part of the state space matrix, in that they do not depend on model parameters. The logical matrix \texttt{mmask} specifies which elements of the stationary \texttt{mat} is dependent on model parameters (variable), while the logical column vector \texttt{dvmask} specifies the variable elements of \texttt{dvec}. Functions that update the state space matrix take the model parameters as input argument, and should output results so that the following code correctly updates the state space matrix:
\begin{center}
\texttt{mat(mmask) = func1(psi);}\\
\texttt{dvec(dvmask, :) = func2(psi);}
\end{center}
where \texttt{func1} update the stationary part, and \texttt{func2} updates the dynamic part. The class \texttt{SSMAT} is designed to represent the ``structure'' and current value of a state space matrix, thus the update functions are not part of the class and will be described in detail in later sections.

Objects of class \texttt{SSMAT} behaves like a matrix to a certain extent, calling \texttt{size} returns the size of the stationary part, concatenations \texttt{horzcat}, \texttt{vertcat} and \texttt{blkdiag} can be used to combine \texttt{SSMAT} with each other and ordinary matrices, and same for \texttt{plus}. Data members \texttt{mat} and \texttt{dvec} can be read and modified provided the structure of the state space matrix stays the same, parenthesis indexing can also be used.

\subsection{Class SSDIST}

The class state space distributions (\texttt{SSDIST}) is a child class of \texttt{SSMAT}, and represents non-Gaussian distributions for the observation and state disturbances ($H_t$ and $Q_t$). Since disturbances are vectors, it's possible to have both elements with Gaussian distributions, and elements with different non-Gaussian distributions. Also in order to use the Kalman filter, non-Gaussian distributions are approximated by Gaussian noise with dynamic variance. Thus the class \texttt{SSDIST} needs to keep track of multiple non-Gaussian distributions and also their corresponding disturbance elements, so that the individual Gaussian approximations can be concatenated in the correct order.

As non-Gaussian distributions are approximated by dynamic Gaussian distribution in SSM, its representation need only consist of data required in the approximation process, namely the functions that take disturbance samples as input and outputs the dynamic Gaussian variance. Also a function that outputs the probability of the disturbance samples is needed for loglikelihood calculation. These functions are stored in two data members \texttt{matf} and \texttt{logpf}, both cell arrays of function handles. A logical matrix \texttt{diagmask} is needed to keep track of which elements each non-Gaussian distribution governs, where each column of \texttt{diagmask} corresponds to each distribution. Elements not specified in any columns are therefore Gaussian. Lastly a logical vector \texttt{dmask} of the same length as \texttt{matf} and \texttt{logpf} specifies which distributions are variable, for example, t-distributions can be variable if the variance and degree of freedom are model parameters. The update function \texttt{func} for \texttt{SSDIST} objects should be defined to take parameter values as input arguments and output a cell matrix of function handles such that
\begin{center}
\tt
distf = func(psi);\\
{[}matf\{dmask\}] = distf\{:, 1\};\\
{[}logpf\{dmask\}] = distf\{:, 2\};
\end{center}
correctly updates the non-Gaussian distributions.

In SSM most non-Gaussian distributions such as exponential family distributions and some heavy-tailed noise distributions are predefined and can be constructed similarly to predefined models, these can then be inserted in stock components to allow for non-Gaussian disturbance generalizations.

\subsection{Class SSFUNC}

The class state space functions (\texttt{SSFUNC}) is another child class of \texttt{SSMAT}, and represents nonlinear functions that transform vectors to vectors ($Z_t$, $T_t$ and $R_t$). As with non-Gaussian distribution it is possible to ``concatenate'' nonlinear functions corresponding to different elements of the state vector and destination vector. The dimension of input and output vector and which elements they corresponds to will have to be kept track of for each nonlinear function, and their linear approximations are concatenated using these information.

Both the function itself and its derivative are needed to calculate the linear approximation, these are stored in data members \texttt{f} and \texttt{df}, both cell arrays of function handles. \texttt{horzmask} and \texttt{vertmask} are both logical matrices where each column specifies the elements of the input and output vector for each function, respectively. In this way the dimensions of the approximating matrix for the \texttt{i}th function can be determined as \texttt{nnz(vertmask(:, i))${}\times{}$nnz(horzmask(:, i))}. The logical vector \texttt{fmask} specifies which functions are variable with respect to model parameters. The update function \texttt{func} for \texttt{SSFUNC} objects should be defined to take parameter values as input arguments and output a cell matrix of function handles such that
\begin{center}
\tt
funcf = func(psi);\\
{[}f\{fmask\}] = funcf\{:, 1\};\\
{[}df\{fmask\}] = funcf\{:, 2\};
\end{center}
correctly updates the nonlinear functions.

\subsection{Class SSPARAM}

The class state space parameters (\texttt{SSPARAM}) stores and organizes the model parameters, each model parameter has a name, value, transform and possible constraints. Transforms are necessary because most model parameters have a restricted domain, but most numerical optimization routines generally operate over the whole real line. For example a variance parameter $\sigma^2$ are suppose to be positive, so if $\sigma^2$ is optimized directly there's the possibility of error, but by defining $\psi=\log(\sigma^2)$ and optimizing $\psi$ instead, this problem is avoided. Model parameter transforms are not restricted to single parameters, sometimes a group of parameters need to be transformed together, the same goes for constraints. Hence the class \texttt{SSPARAM} also keeps track of parameter grouping.

Like \texttt{SSMAT}, \texttt{SSPARAM} objects can be treated like a row vector of parameters to a certain extent. Specifically, it is possible to horizontally concatenate \texttt{SSPARAM} objects as a means of combining parameter groups. Other operations on \texttt{SSPARAM} objects include getting and setting the parameter values and transformed values. The untransformed values are the original parameter values meaningful to the model defined, and will be referred to as \texttt{param} in code examples; the transformed values are meant only as input to optimization routines and have no direct interpretation in terms of the model, they are referred to as \texttt{psi} in code examples, but the user will rarely need to use them directly if predefined models are used.

\subsection{Class SSMODEL}

The class state space model (\texttt{SSMODEL}) represents state space models by embedding \texttt{SSMAT}, \texttt{SSDIST}, \texttt{SSFUNC} and \texttt{SSPARAM} objects and also keeping track of update functions, parameter masks and general model component information. The basic constituent elements of a model is described earlier as \texttt{Z}, \texttt{T}, \texttt{R}, \texttt{H}, \texttt{Q}, \texttt{c}, \texttt{a1} and \texttt{P1}. \texttt{Z}, \texttt{T} and \texttt{R} are vector transformations, the first two can be \texttt{SSFUNC} or \texttt{SSMAT} objects, but the last one can only be a \texttt{SSMAT} object. \texttt{H} and \texttt{Q} govern the disturbances and can be \texttt{SSDIST} or \texttt{SSMAT} objects. \texttt{c}, \texttt{a1} and \texttt{P1} can only be \texttt{SSMAT} objects. These form the ``constant'' part of the model specification, and is the only part needed in Kalman filter and related state estimation algorithms.

The second part of \texttt{SSMODEL} concerns the update functions and model parameters, general specifications for the model estimation process. \texttt{func} and \texttt{grad} are cell arrays of update functions and their gradient for the basic model elements respectively. It is possible for a single function to update multiple parts of the model and these information are stored in the data member \texttt{A}, a \texttt{length(func)}${}\times{}$\texttt{18} adjacency matrix. Each row of \texttt{A} represents one update function, and each column represents one updatable element. The \texttt{18} updatable model elements in order are
\begin{center}
\texttt{H Z T R Q c a1 P1 Hd Zd Td Rd Qd cd Hng Qng Znl Tnl}
\end{center}
where the \texttt{d} suffix means dynamic, \texttt{ng} means non-Gaussian and \texttt{nl} means nonlinear. It is therefore possible for a function that updates \texttt{Z} to output \texttt{[vec dsubvec funcf]} updating the stationary part of \texttt{Z}, dynamic part of \texttt{Z} and the nonlinear functions in \texttt{Z} respectively. The row in \texttt{A} for this function would be
\begin{center}
\begin{tt}
\begin{tabular}{*{17}{r@{ }}r}
H & Z & T & R & Q & c & a1 & P1 & Hd & Zd & Td & Rd & Qd & cd & Hng & Qng & Znl & Tnl \\
0 & 1 & 0 & 0 & 0 & 0 &  0 &  0 &  0 &  1 &  0 &  0 &  0 &  0 &   0 &   0 &   1 &   0
\end{tabular}
\end{tt}
\end{center}
Any function can update any combination of these 18 elements, the only restriction is that the output should follow the order set in \texttt{A} with possible omissions.

On the other hand \texttt{A} also allows many functions to update a single model element, this is the main mechanism that enables smooth model combination. Suppose \texttt{T1} of model $1$ is updated by \texttt{func1}, and \texttt{T2} of model $2$ by \texttt{func2}, combination of models $1$ and $2$ requires block diagonal concatenation of \texttt{T1} and \texttt{T2}: \texttt{T = blkdiag(T1, T2)}. It is easy to see that the vertical concatenation of the output of \texttt{func1} and \texttt{func2} correctly updates the new matrix \texttt{T} as follows
\begin{center}
\tt
vec1 = func1(psi);\\
vec2 = func2(psi);\\
T.mat(T.mmask) = [vec1; vec2];
\end{center}
This also holds for horizontal concatenation, but for vertical concatenation of more than one column this would fail, luckily this case never occurs when combining state space models, so can be safely ignored. The updates of dynamic, non-Gaussian and nonlinear elements can also be combined in this way, in fact, the whole scheme of the update functions are designed with combinable output in mind. For an example adjacency matrix \texttt{A}
\begin{center}
\tt
\begin{tabular}{l*{17}{r@{ }}r}
      & H & Z & T & R & Q & c & a1 & P1 & Hd & Zd & Td & Rd & Qd & cd & Hng & Qng & Znl & Tnl \\
func1 & 1 & 0 & 1 & 0 & 1 & 0 &  0 &  0 &  0 &  0 &  0 &  0 &  0 &  0 &   0 &   0 &   1 &   0 \\
func2 & 0 & 0 & 1 & 0 & 0 & 0 &  0 &  0 &  0 &  1 &  0 &  0 &  0 &  0 &   0 &   0 &   0 &   0 \\
func3 & 0 & 1 & 1 & 0 & 1 & 0 &  0 &  0 &  0 &  1 &  0 &  0 &  0 &  0 &   0 &   0 &   1 &   0
\end{tabular}
\end{center}
model update would then proceed by first invoking the three update functions
\begin{center}
\tt
[Hvec1 Tvec1 Qvec1 Zfuncf1] = func1(psi);\\
{[}Tvec2 Zdsubvec2] = func2(psi);\\
{[}Zvec3 Tvec3 Qvec3 Zdsubvec3 Zfuncf3] = func3(psi);
\end{center}
and then update the various elements
\begin{center}
\tt
H.mat(H.mmask) = Hvec1;\\
Z.mat(Z.mmask) = Zvec3;\\
T.mat(T.mmask) = [Tvec1; Tvec2; Tvec3];\\
Q.mat(Q.mmask) = [Qvec1; Qvec3];\\
Z.dvec(Z.dvmask, :) = [Zdsubvec2; Zdsubvec3];\\
Z.f(Z.fmask) = [Zfuncf1\{:, 1\}; Zfuncf3\{:, 1\}];\\
Z.df(Z.fmask) = [Zfuncf1\{:, 2\}; Zfuncf3\{:, 2\}];
\end{center}

Now to facilitate model construction and alteration, functions that need adjacency matrix \texttt{A} as input actually expects another form based on strings. Each of the 18 elements shown earlier is represented by the corresponding strings, \texttt{'H'}, \texttt{'Qng'}, \texttt{'Zd'}, \ldots etc., and each row of \texttt{A} is a single string in a cell array of strings, where each string is a concatenation of various elements in any order. This form (a cell array of strings) is called ``adjacency string'' and is used in all functions that require an adjacency matrix as input.

The remaining data members of \texttt{SSMODEL} are \texttt{psi} and \texttt{pmask}. \texttt{psi} is a \texttt{SSPARAM} object that stores the model parameters, and \texttt{pmask} is a cell array of logical row vectors that specifies which parameters are required by each update function. So each update function are in fact provided a subset of model parameters \texttt{func$i$(psi.value(pmask\{$i$\}))}. It is trivial to combine these two data members when combining models.

In summary the class \texttt{SSMODEL} can be divided into two parts, a ``constant'' part that corresponds to an instance of a theoretical state space model given all model parameters, which is the only part required to run Kalman filter and related state estimation routines, and a ``variable'' part that keeps track of the model parameters and how each parameter effects the values of various model elements, which is used (in addition to the ``constant'' part) in model estimation routines that produce parameter estimates. This division is the careful separation of the ``structure'' of a model and its ``instantiation'' given model parameters, and allows logical separation of various structural elements of the model (with the classes \texttt{SSMAT}, \texttt{SSDIST} and \texttt{SSFUNC}) to facilitate model combination and usage, without compromising the ability to efficiently define and combine complex model update mechanisms.

\section{SSM Functions}

There are several types of functions in SSM besides class methods, these are data analysis functions, stock element functions and various helper functions. Most data analysis functions are Kalman filter related and implemented as \texttt{SSMODEL} methods, but for the user this is transparent since class methods and functions are called in the same way in MATLAB. Examples for data analysis functions include \texttt{kalman}, \texttt{statesmo} and \texttt{loglik}, all of which uses the kalman filter and outputs the filtered states, smoothed states and loglikelihood respectively. These and similar functions take three arguments, the data \texttt{y}, which must be \texttt{p}${}\times{}$\texttt{n}, where \texttt{p} is the number of variables in the data and \texttt{n} the data size, the model \texttt{model}, a \texttt{SSMODEL} object, and optional analysis option name value pairs (see function \texttt{setopt} in Appendix \ref{app:function:helper}), where settings such as tolerance can be specified. The functions \texttt{linear} and \texttt{gauss} calculates the linear Gaussian approximation models, and requires \texttt{y}, \texttt{model} and optional initial state estimate \texttt{alpha0} as input arguments. The function \texttt{estimate} runs numerical optimization routines for model estimation; model updates, necessary approximations and loglikelihood calculations are done automatically in the procedure, input arguments are \texttt{y}, \texttt{model} and initial parameter estimate \texttt{psi0}, the function outputs the maximum likelihood model. There are also some functions specific to ARIMA-type models, such as Hillmer-Tiao decomposition \cite{Hillmer82} and TRAMO model selection \cite{Gomez01}.

Stock element functions create individual parts of predefined models, they range from the general to the specific. These functions have names of the form \texttt{(element)\_(description)}, for example, the function \texttt{mat\_arma} creates matrices for the ARMA model. Currently there are five types of stock element functions: \texttt{fun\_*} creates stationary or dynamic matrix update functions, \texttt{mat\_*} creates matrix structures, \texttt{ngdist\_*} creates non-Gaussian distributions (or the SSM representation of which), \texttt{ngfun\_*} creates non-Gaussian distribution update functions, and \texttt{x\_*} creates stock regression variables such as trading day variables. Note that these functions are created to be class independent whenever possible, in that they do not directly construct or output SSM class objects, although the form of the output may \emph{imply} the SSM class structures (an exception is that \texttt{fun\_*} functions return a \texttt{SSPARAM} object). This is primarily due to modular considerations and also that these ``utility'' functions can then be useful to users who (for whatever reason) do not wish to use the SSM classes.

Miscellaneous helper functions provide additional functionality unsuitable to be implemented in the previous two categories. The most important of which is the function \texttt{setopt}, which specifies the global options or settings structure that is used by almost every data analysis function. One time overrides of individual options are possible for each function call, and each option can be individually specified, examples of possible options include tolerance and verbosity. Other useful functions include \texttt{ymarray} which generates a year, month array from starting year, starting month and number of months and is useful for functions like \texttt{x\_td} and \texttt{x\_ee} which take year, month arrays as input, \texttt{randarma} which generates a random AR or MA polynomial, and \texttt{meancov} which calculates the mean and covariance of a vector sequence.

\section{Data Analysis Examples}

Many of the data analysis examples are based on the data in the book by Durbin and Koopman \cite{Durbin01}, to facilitate easy comparison and understanding of the method used here.

\subsection{Structural time series models}
\label{sec:seatbelt}

In this example data on road accidents in Great Britain \cite{Harvey86} is analyzed using structural time series models following \cite{Durbin01}. The purpose the the analysis is to assess the effect of seat belt laws on road accident casualties, with individual monthly figures for drivers, front seat passengers and rear seat passengers. The monthly price of petrol and average number of kilometers traveled will be used as regression variables. The data is from January 1969 to December 1984.

\subsubsection{Univariate analysis}

The drivers series will be analyzed with univariate structural time series model, which consists of local level component $\mu_t$, trigonometric seasonal component $\gamma_t$, regression component (on price of petrol) and intervention component (introduction of seat belt law) $\beta x_t$. The model equation is
\begin{equation}
y_t=\mu_t+\gamma_t+\beta x_t+\varepsilon_t,
\end{equation}
where $\varepsilon_t$ is the observation noise. The following code example constructs this model:
\begin{small}\begin{verbatim}
lvl = ssmodel('llm');
seas = ssmodel('seasonal', 'trig1', 12);
intv = ssmodel('intv', n, 'step', 170);
reg = ssmodel('reg', petrol, 'price of petrol');
bstsm = [lvl seas intv reg];
bstsm.name = 'Basic structural time series model';
bstsm.param = [10 0.1 0.001];
\end{verbatim}\end{small}
The MATLAB display for object \texttt{bstsm} is
\begin{small}\begin{verbatim}
bstsm =

    Basic structural time series model
    ==================================

    p = 1
    m = 14
    r = 12
    n = 192


    Components (5):
    ==================

    [Gaussian noise]
        p               = 1
    [local polynomial trend]
        d               = 0
    [seasonal]
        subtype         = trig1
        s               = 12
    [intervention]
        subtype         = step
        tau             = 170
    [regression]
        variable        = price of petrol


    H matrix (1, 1)
    -------------------

      10.0000


    Z matrix (1, 14, 192)
    ----------------------

    Stationary part:

        1    1    0    1    0    1    0    1    0    1    0    1  DYN  DYN


    Dynamic part:

    Columns 1 through 9

            0        0        0        0        0        0        0        0        0
      -2.2733  -2.2792  -2.2822  -2.2939  -2.2924  -2.2968  -2.2655  -2.2626  -2.2655

    Columns 10 through 18

            0        0        0        0        0        0        0        0        0
      -2.2728  -2.2757  -2.2828  -2.2899  -2.2956  -2.3012  -2.3165  -2.3192  -2.3220


                                          ...

    Columns 181 through 189

            1        1        1        1        1        1        1        1        1
      -2.1390  -2.1646  -2.1565  -2.1597  -2.1644  -2.1648  -2.1634  -2.1646  -2.1707

    Columns 190 through 192

            1        1        1
      -2.1502  -2.1539  -2.1536


    T matrix (14, 14)
    -------------------

    Columns 1 through 9

            1        0        0        0        0        0        0        0        0
            0   0.8660   0.5000        0        0        0        0        0        0
            0  -0.5000   0.8660        0        0        0        0        0        0
            0        0        0   0.5000   0.8660        0        0        0        0
            0        0        0  -0.8660   0.5000        0        0        0        0
            0        0        0        0        0   0.0000        1        0        0
            0        0        0        0        0       -1   0.0000        0        0
            0        0        0        0        0        0        0  -0.5000   0.8660
            0        0        0        0        0        0        0  -0.8660  -0.5000
            0        0        0        0        0        0        0        0        0
            0        0        0        0        0        0        0        0        0
            0        0        0        0        0        0        0        0        0
            0        0        0        0        0        0        0        0        0
            0        0        0        0        0        0        0        0        0

    Columns 10 through 14

            0        0        0        0        0
            0        0        0        0        0
            0        0        0        0        0
            0        0        0        0        0
            0        0        0        0        0
            0        0        0        0        0
            0        0        0        0        0
            0        0        0        0        0
            0        0        0        0        0
      -0.8660   0.5000        0        0        0
      -0.5000  -0.8660        0        0        0
            0        0       -1        0        0
            0        0        0        1        0
            0        0        0        0        1


    R matrix (14, 12)
    -------------------

        1    0    0    0    0    0    0    0    0    0    0    0
        0    1    0    0    0    0    0    0    0    0    0    0
        0    0    1    0    0    0    0    0    0    0    0    0
        0    0    0    1    0    0    0    0    0    0    0    0
        0    0    0    0    1    0    0    0    0    0    0    0
        0    0    0    0    0    1    0    0    0    0    0    0
        0    0    0    0    0    0    1    0    0    0    0    0
        0    0    0    0    0    0    0    1    0    0    0    0
        0    0    0    0    0    0    0    0    1    0    0    0
        0    0    0    0    0    0    0    0    0    1    0    0
        0    0    0    0    0    0    0    0    0    0    1    0
        0    0    0    0    0    0    0    0    0    0    0    1
        0    0    0    0    0    0    0    0    0    0    0    0
        0    0    0    0    0    0    0    0    0    0    0    0


    Q matrix (12, 12)
    -------------------

    Columns 1 through 9

       0.1000        0        0        0        0        0        0        0        0
            0   0.0010        0        0        0        0        0        0        0
            0        0   0.0010        0        0        0        0        0        0
            0        0        0   0.0010        0        0        0        0        0
            0        0        0        0   0.0010        0        0        0        0
            0        0        0        0        0   0.0010        0        0        0
            0        0        0        0        0        0   0.0010        0        0
            0        0        0        0        0        0        0   0.0010        0
            0        0        0        0        0        0        0        0   0.0010
            0        0        0        0        0        0        0        0        0
            0        0        0        0        0        0        0        0        0
            0        0        0        0        0        0        0        0        0

    Columns 10 through 12

            0        0        0
            0        0        0
            0        0        0
            0        0        0
            0        0        0
            0        0        0
            0        0        0
            0        0        0
            0        0        0
       0.0010        0        0
            0   0.0010        0
            0        0   0.0010


    P1 matrix (14, 14)
    -------------------

      Inf    0    0    0    0    0    0    0    0    0    0    0    0    0
        0  Inf    0    0    0    0    0    0    0    0    0    0    0    0
        0    0  Inf    0    0    0    0    0    0    0    0    0    0    0
        0    0    0  Inf    0    0    0    0    0    0    0    0    0    0
        0    0    0    0  Inf    0    0    0    0    0    0    0    0    0
        0    0    0    0    0  Inf    0    0    0    0    0    0    0    0
        0    0    0    0    0    0  Inf    0    0    0    0    0    0    0
        0    0    0    0    0    0    0  Inf    0    0    0    0    0    0
        0    0    0    0    0    0    0    0  Inf    0    0    0    0    0
        0    0    0    0    0    0    0    0    0  Inf    0    0    0    0
        0    0    0    0    0    0    0    0    0    0  Inf    0    0    0
        0    0    0    0    0    0    0    0    0    0    0  Inf    0    0
        0    0    0    0    0    0    0    0    0    0    0    0  Inf    0
        0    0    0    0    0    0    0    0    0    0    0    0    0  Inf


    Adjacency matrix (3 functions):
    -----------------------------------------------------------------------------
    Function                H Z T R Q c a1 P1 Hd Zd Td Rd Qd cd Hng Qng Znl Tnl
    fun_var/psi2var         1 0 0 0 0 0 0  0  0  0  0  0  0  0  0   0   0   0
    fun_var/psi2var         0 0 0 0 1 0 0  0  0  0  0  0  0  0  0   0   0   0
    fun_dupvar/psi2dupvar1  0 0 0 0 1 0 0  0  0  0  0  0  0  0  0   0   0   0


    Parameters (3)
    ------------------------------------
    Name                Value
    epsilon var         10
    zeta var            0.1
    omega var           0.001
\end{verbatim}\end{small}
The constituting components of the model and the model matrices are all displayed, note that \texttt{Z} is dynamic with the intervention and regression variables, and matrices \texttt{c} and \texttt{a1} are omitted if they are zero, also \texttt{H} and \texttt{Q} take on values determined by the parameters set.

\begin{figure}
\includegraphics[width=1\textwidth]{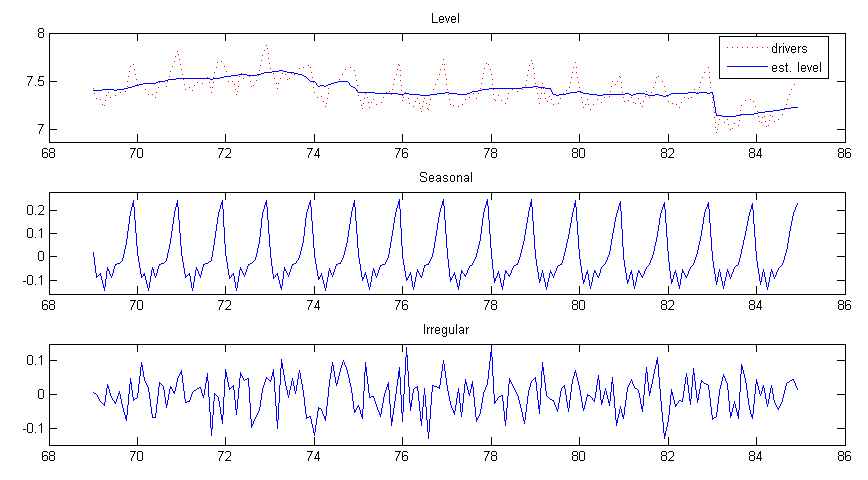}
\caption{Driver casualties estimated by basic structural time series model}
\label{fig:seatbelt}
\end{figure}

With the model constructed, estimation can proceed with the code:
\begin{small}\begin{verbatim}
[bstsm logL] = estimate(y, bstsm);
[alphahat V] = statesmo(y, bstsm);
irr = disturbsmo(y, bstsm);
ycom = signal(alphahat, bstsm);
ylvl = sum(ycom([1 3 4], :), 1);
yseas = ycom(2, :);
\end{verbatim}\end{small}
The estimated model parameters are \texttt{[0.0037862 0.00026768 1.162e-006]}, which can be obtained by displaying \texttt{bstsm.param}, the loglikelihood is \texttt{175.7790}. State and disturbance smoothing is performed with the estimated model, and the smoothed state is transformed into signal components by \texttt{signal}. Because \texttt{p}${}=1$, the output \texttt{ycom} is \texttt{M}${}\times{}$\texttt{n}, where \texttt{M} is the number of signal components. The level, intervention and regression are summed as the total data level, separating seasonal influence. Using MATLAB graphic functions the individual signal components and data are visualized:
\begin{small}\begin{verbatim}
subplot(3, 1, 1), plot(time, y, 'r:', 'DisplayName', 'drivers'), hold all,
                  plot(time, ylvl, 'DisplayName', 'est. level'), hold off,
                  title('Level'), ylim([6.875 8]), legend('show');
subplot(3, 1, 2), plot(time, yseas), title('Seasonal'), ylim([-0.16 0.28]);
subplot(3, 1, 3), plot(time, irr), title('Irregular'), ylim([-0.15 0.15]);
\end{verbatim}\end{small}
The graphical output is shown in figure \ref{fig:seatbelt}.
The coefficient for the intervention and regression is defined as part of the state vector in this model, so they can be obtained from the last two elements of the smoothed state vector (due to the order of component concatenation) at the last time point. Coefficient for the intervention is \texttt{alphahat(13, end) = -0.23773}, and coefficient for the regression (price of petrol) is \texttt{alphahat(14, end) = -0.2914}. In this way diagnostics of these coefficient estimates can also be obtained by the smoothed state variance \texttt{V}.

\subsubsection{Bivariate analysis}

\begin{figure}
\includegraphics[width=1\textwidth]{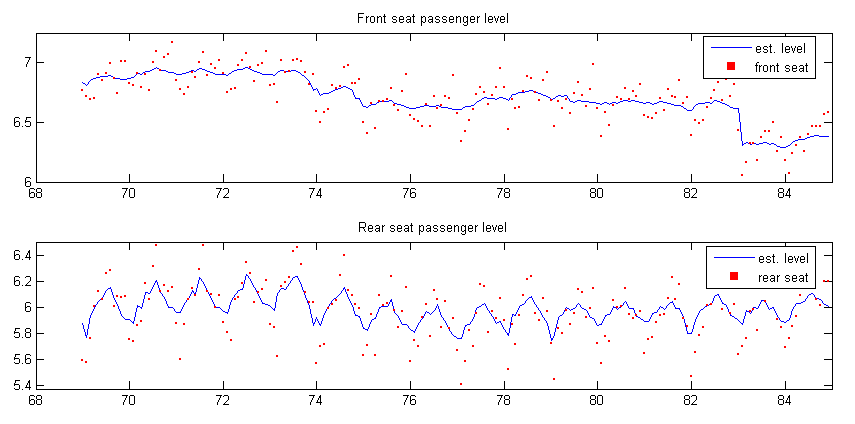}
\caption{Passenger casualties estimated by bivariate structural time series model}
\label{fig:seatbelt2}
\end{figure}

The front seat passenger and rear seat passenger series will be analyzed together using bivariate structural time series model, with components as before. Specifically, separate level and seasonal components are defined for both series, but the disturbances are assumed to be correlated. To reduce the number of parameters estimated the seasonal component is assumed to be fixed, so that the total number of parameters is six. We also include regression on the price of petrol and kilometers traveled, and intervention for only the first series, since the seat belt law only effects the front seat passengers. The following is the model construction code:
\begin{small}\begin{verbatim}
bilvl = ssmodel('mvllm', 2);
biseas = ssmodel('mvseasonal', 2, [], 'trig fixed', 12);
biintv = ssmodel('mvintv', 2, n, {'step' 'null'}, 170);
bireg = ssmodel('mvreg', 2, [petrol; km]);
bistsm = [bilvl biseas biintv bireg];
bistsm.name = 'Bivariate structural time series model';
\end{verbatim}\end{small}
The model is then estimated with carefully chosen initial values, and state smoothing and signal extraction proceeds as before:
\begin{small}\begin{verbatim}
[bistsm logL] = estimate(y2, bistsm, [0.0054 0.0086 0.0045 0.00027 0.00024 0.00023]);
[alphahat V] = statesmo(y2, bistsm);
y2com = signal(alphahat, bistsm);
y2lvl = sum(y2com(:,:, [1 3 4]), 3);
y2seas = y2com(:,:, 2);
\end{verbatim}\end{small}
When \texttt{p}${}>1$ the output from \texttt{signal} is of dimension \texttt{p}${}\times{}$\texttt{n}${}\times{}$\texttt{M}, where \texttt{M} is the number of components. The level, regression and intervention are treated as one component of data level as before, separated from the seasonal component. The components estimated for the two series is plotted:
\begin{footnotesize}\begin{verbatim}
subplot(2, 1, 1), plot(time, y2lvl(1, :), 'DisplayName', 'est. level'), hold all,
                  scatter(time, y2(1, :), 8, 'r', 's', 'filled', 'DisplayName', 'front seat'), hold off,
                  title('Front seat passenger level'), xlim([68 85]), ylim([6 7.25]), legend('show');
subplot(2, 1, 2), plot(time, y2lvl(2, :), 'DisplayName', 'est. level'), hold all,
                  scatter(time, y2(2, :), 8, 'r', 's', 'filled', 'DisplayName', 'rear seat'), hold off,
                  title('Rear seat passenger level'), xlim([68 85]), ylim([5.375 6.5]), legend('show');
\end{verbatim}\end{footnotesize}
The graphical output is shown in figure \ref{fig:seatbelt2}.
The intervention coefficient for the front passenger series is obtained by \texttt{alphahat(25, end) = -0.30025}, the next four elements are the coefficients of the regression of the two series on the price of petrol and kilometers traveled.

\subsection{ARMA models}

In this example the difference of the number of users logged on an internet server \cite{Makridakis98} is analyzed by ARMA models, and model selection via BIC and missing data analysis are demonstrated. To select an appropriate ARMA$(p, q)$ model for the data various values for $p$ and $q$ are tried, and the BIC of the estimated model for each is recorded, the model with the lowest BIC value is chosen.
\begin{small}\begin{verbatim}
for p = 0 : 5
    for q = 0 : 5
        arma = ssmodel('arma', p, q);
        [arma logL output] = estimate(y, arma, 0.1);
        BIC(p+1, q+1) = output.BIC;
    end
end
[m i] = min(BIC(:));
arma = estimate(y, ssmodel('arma', mod(i-1, 6), floor((i-1)/6)), 0.1);
\end{verbatim}\end{small}
The BIC values obtained for each model is as follows:\medskip\\
\begin{tabular}{lrrrrrr}
      & $q=0$  & $q=1$  & $q=2$  & $q=3$  & $q=4$  & $q=5$  \\
$p=0$ & 6.3999 & 5.6060 & 5.3299 & 5.3601 & 5.4189 & 5.3984 \\
$p=1$ & 5.3983 & 5.2736 & 5.3195 & 5.3288 & 5.3603 & 5.3985 \\
$p=2$ & 5.3532 & 5.3199 & 5.3629 & 5.3675 & 5.3970 & 5.4436 \\
$p=3$ & 5.2765 & 5.3224 & 5.3714 & 5.4166 & 5.4525 & 5.4909 \\
$p=4$ & 5.3223 & 5.3692 & 5.4142 & 5.4539 & 5.4805 & 5.4915 \\
$p=5$ & 5.3689 & 5.4124 & 5.4617 & 5.5288 & 5.5364 & 5.5871
\end{tabular}\medskip\\
The model with the lowest BIC is ARMA$(1, 1)$, second lowest is ARMA$(3, 0)$, the former model is chosen for subsequent analysis.

Next missing data is simulated by setting some time points to \texttt{NaN}, model and state estimation can still proceed normally with missing data present.
\begin{small}\begin{verbatim}
y([6 16 26 36 46 56 66 72 73 74 75 76 86 96]) = NaN;
arma = estimate(y, arma, 0.1);
yf = signal(kalman(y, arma), arma);
\end{verbatim}\end{small}

\begin{figure}
\includegraphics[width=1\textwidth]{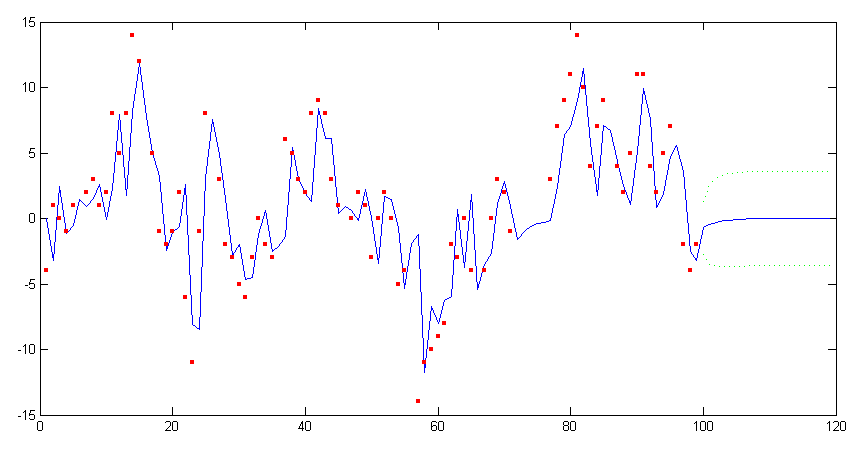}
\caption{Forecast using ARMA$(1, 1)$ model with 50\% confidence interval}
\label{fig:internet}
\end{figure}

Forecasting is equivalent to treating future data as missing, thus the data set \texttt{y} is appended with as many \texttt{NaN} values as the steps ahead to forecast. Using the previous estimated ARMA$(1,1)$ model the Kalman filter will then effectively predict future data points.
\begin{small}\begin{verbatim}
[a P v F] = kalman([y repmat(NaN, 1, 20)], arma);
yf = signal(a(:, 1:end-1), arma);
conf50 = 0.675*realsqrt(F); conf50(1:n) = NaN;
plot(yf), hold all, plot([yf+conf50; yf-conf50]', 'g:'),
scatter(1:length(y), y, 10, 'r', 's', 'filled'), hold off, ylim([-15 15]);
\end{verbatim}\end{small}
The plot of the forecast and its confidence interval is shown in figure \ref{fig:internet}. Note that if the Kalman filter is replaced with the state smoother, the forecasted values will still be the same.

\subsection{Cubic spline smoothing}

\begin{figure}
\includegraphics[width=1\textwidth]{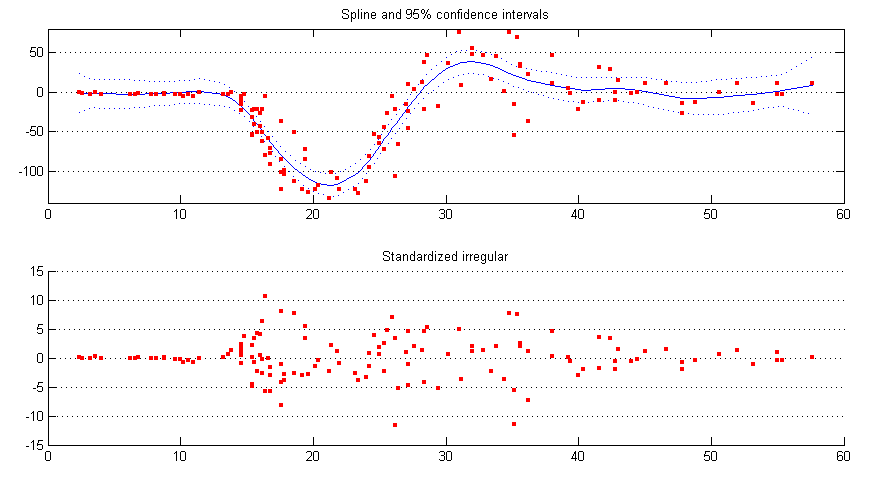}
\caption{Cubic spline smoothing of motorcycle acceleration data}
\label{fig:motorcycle}
\end{figure}

The general state space formulation can also be used to do cubic spline smoothing, by putting the cubic spline into an equivalent state space form, and accounting for the continuous nature of such smoothing procedures. Here the continuous acceleration data of a simulated motorcycle accident \cite{Silverman85} is smoothed by the cubic spline model, which is predefined.
\begin{small}\begin{verbatim}
spline = estimate(y, ssmodel('spline', delta), [1 0.1]);
[alphahat V] = statesmo(y, spline);
conf95 = squeeze(1.96*realsqrt(V(1, 1, :)))';
[eps eta epsvar] = disturbsmo(y, spline);
\end{verbatim}\end{small}
The smoothed data and standardized irregular is plotted and shown in figure \ref{fig:motorcycle}.
\begin{small}\begin{verbatim}
subplot(2, 1, 1), plot(time, alphahat(1, :), 'b'), hold all,
                  plot(time, [alphahat(1, :) + conf95; alphahat(1, :) - conf95], 'b:'),
                  scatter(time, y, 10, 'r', 's', 'filled'), hold off,
                  title('Spline and 95% confidence intervals'), ylim([-140 80]),
                  set(gca,'YGrid','on');
subplot(2, 1, 2), scatter(time, eps./realsqrt(epsvar), 10, 'r', 's', 'filled'),
                  title('Standardized irregular'), set(gca,'YGrid','on');
\end{verbatim}\end{small}

\begin{figure}
\includegraphics[width=1\textwidth]{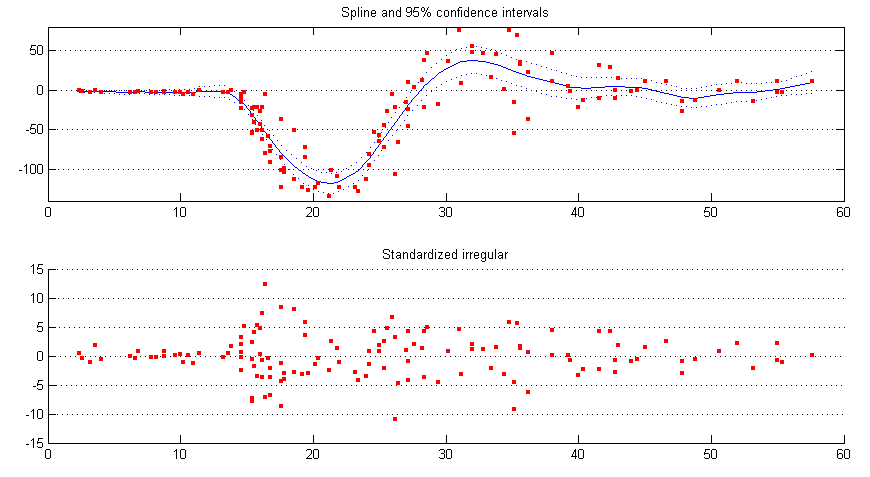}
\caption{Cubic spline smoothing of motorcycle acceleration data with heteroscedastic noise}
\label{fig:motorcycle2}
\end{figure}

It is seen from figure \ref{fig:motorcycle} that the irregular may be heteroscedastic, an easy \textit{ad hoc} solution is to model the changing variance of the irregular by a continuous version of the local level model. Assume the irregular variance is $\sigma^2_\varepsilon h_t^2$ at time point $t$ and $h_1=1$, then we model the absolute value of the smoothed irregular \texttt{abs(eps)} with $h_t$ as its level. The continuous local level model with level $h_t$ needs to be constructed from scratch.
\begin{small}\begin{verbatim}
contllm = ssmodel('', 'continuous local level', ...
                  0, 1, 1, 1, ssmat(0, [], true, zeros(1, n), true), ...
                  'Qd', {@(X) exp(2*X)*delta}, {[]}, ssparam({'zeta var'}, '1/2 log'));
contllm = [ssmodel('Gaussian') contllm];
alphahat = statesmo(abs(eps), estimate(abs(eps), contllm, [1 0.1]));
h2 = (alphahat/alphahat(1)).^2;
\end{verbatim}\end{small}
\texttt{h2} is then the relative magnitude of the noise variance $h_t^2$ at each time point, which can be used to construct a custom dynamic observation noise model as follows.
\begin{small}\begin{verbatim}
hetnoise = ssmodel('', 'Heteroscedastic noise', ...
                   ssmat(0, [], true, zeros(1, n), true), zeros(1, 0), [], [], [], ...
                   'Hd', {@(X) exp(2*X)*h2}, {[]}, ssparam({'epsilon var'}, '1/2 log'));
hetspline = estimate(y, [hetnoise spline], [1 0.1]);
[alphahat V] = statesmo(y, hetspline);
conf95 = squeeze(1.96*realsqrt(V(1, 1, :)))';
[eps eta epsvar] = disturbsmo(y, hetspline);
\end{verbatim}\end{small}
The smoothed data and standardized irregular with heteroscedastic assumption is shown in figure \ref{fig:motorcycle2}, it is seen that the confidence interval shrank, especially at the start and end of the series, and the irregular is slightly more uniform.

In summary the motorcycle series is first modeled by a cubic spline model with Gaussian irregular assumption, then the smoothed irregular magnitude itself is modeled with a local level model. Using the irregular level estimated at each time point as the relative scale of irregular variance, a new heteroscedastic irregular continuous model is constructed with the estimated level built-in, and plugged into the cubic spline model to obtain new estimates for the motorcycle series.

\subsection{Poisson distribution error model}

The road accident casualties and seat belt law data analyzed in section \ref{sec:seatbelt} also contains a monthly van driver casualties series. Due to the smaller numbers of van driver casualties the Gaussian assumption is not justified in this case, previous methods cannot be applied. Here a poisson distribution is assumed for the data, the mean is $\exp(\theta_t)$ and the observation distribution is
\[p(y_t|\theta_t)=\exp\left(\theta_t^Ty_t-\exp(\theta_t)-\log y_t!\right)\]
The signal $\theta_t$ in turn is modeled with the structural time series model. The total model is then constructed by concatenating a poisson distribution model with a standard structural time series model, the former model will replace the default Gaussian noise model.
\begin{small}\begin{verbatim}
pbstsm = [ssmodel('poisson') ssmodel('llm') ...
          ssmodel('seasonal', 'dummy fixed', 12) ssmodel('intv', n, 'step', 170)];
pbstsm.name = 'Poisson basic STSM';
\end{verbatim}\end{small}
Model estimation will automatically calculate the Gaussian approximation to the poisson model. Since this is an exponential family distribution the data $y_t$ also need to be transformed to $\tilde{y}_t$, which is stored in the output argument \texttt{output}, and used in place of $y_t$ for all functions implementing linear Gaussian (Kalman filter related) algorithms. The following is the code for model and state estimation
\begin{small}\begin{verbatim}
[pbstsm logL output] = estimate(y, pbstsm, 0.0006, [], 'fmin', 'bfgs', 'disp', 'iter');
[alphahat V] = statesmo(output.ytilde, pbstsm);
thetacom = signal(alphahat, pbstsm);
\end{verbatim}\end{small}
Note that the original data \texttt{y} is input to the model estimation routine, which also calculates the transform \texttt{ytilde}. The model estimated then has its Gaussian approximation built-in, and will be treated by the state smoother as a linear Gaussian model, hence the transformed data \texttt{ytilde} needs to be used as input. The signal components \texttt{thetacom} obtained from the smoothed state \texttt{alphahat} is the components of $\theta_t$, the mean of \texttt{y} can then be estimated by \texttt{exp(sum(thetacom, 1))}.

\begin{figure}
\includegraphics[width=1\textwidth]{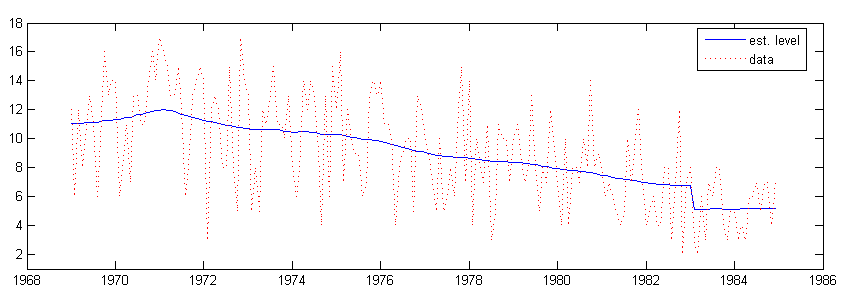}
\caption{Van driver casualties and estimated level}
\label{fig:van}
\end{figure}

The exponentiated level $\exp(\theta_t)$ with the seasonal component eliminated is compared to the original data in figure \ref{fig:van}. The effect of the seat belt law can be clearly seen.
\begin{small}\begin{verbatim}
thetalvl = thetacom(1, :) + thetacom(3, :);
plot(time, exp(thetalvl), 'DisplayName', 'est. level'), hold all,
plot(time, y, 'r:', 'DisplayName', 'data'), hold off, ylim([1 18]), legend('show');
\end{verbatim}\end{small}

\subsection{t-distribution models}

\begin{figure}
\includegraphics[width=1\textwidth]{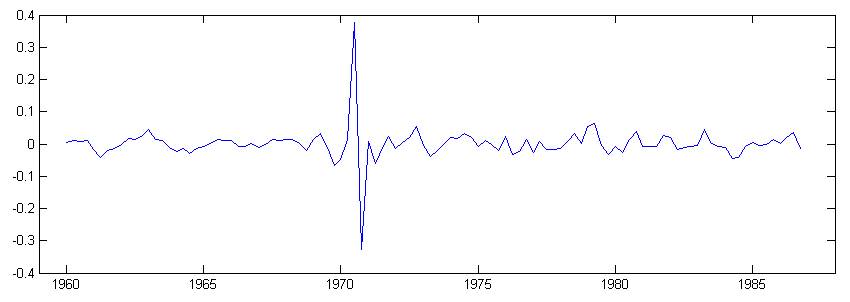}
\caption{t-distribution irregular}
\label{fig:gas}
\end{figure}

In this example another kind of non-Gaussian models, t-distribution models is used to analyze quarterly demand for gas in UK \cite{Koopman00}. A structural time series model with a t-distribution heavy-tailed observation noise is constructed similar to the last section, and model estimation and state smoothing performed.
\begin{small}\begin{verbatim}
tstsm = [ssmodel('t') ssmodel('llt') ssmodel('seasonal', 'dummy', 4)];
tstsm = estimate(y, tstsm, [0.0018 4 7.7e-10 7.9e-6 0.0033], [], 'fmin', 'bfgs', 'disp', 'iter');
[alpha irr] = fastsmo(y, tstsm);
plot(time, irr), title('Irregular component'), xlim([1959 1988]), ylim([-0.4 0.4]);
\end{verbatim}\end{small}
Since t-distribution is not an exponential family distribution, data transformation is not necessary, and \texttt{y} is used throughout. A plot of the irregular in figure \ref{fig:gas} shows that potential outliers with respect to the Gaussian assumption has been detected by the use of heavy-tailed distribution.

\subsection{Hillmer-Tiao decomposition}

In this example seasonal adjustment is performed by the Hillmer-Tiao decomposition \cite{Hillmer82} of airline models. The data (Manufacturing and reproducing magnetic and optical media, U.S. Census Bureau) is fitted with the airline model, then the estimated model is Hillmer-Tiao decomposed into an ARIMA components model with trend and seasonal components. The same is done for the generalized airline model\footnote{Currently renamed Frequency Specific ARIMA model.} \cite{Aston07}, and the seasonal adjustment results are compared. The following are the code to perform seasonal adjustment with the airline model:
\begin{small}\begin{verbatim}
air = estimate(y, ssmodel('airline'), 0.1);
aircom = ssmhtd(air);
ycom = signal(statesmo(y, aircom), aircom);
airseas = ycom(2, :);
\end{verbatim}\end{small}
\texttt{aircom} is the decomposed ARIMA components model corresponding to the estimated airline model, and \texttt{airseas} is the seasonal component, which will be subtracted out of the data \texttt{y} to obtain the seasonal adjusted series. \texttt{ssmhtd} automatically decompose ARIMA type models into trend, seasonal and irregular components, plus any extra MA components as permissible.

\begin{figure}
\includegraphics[width=1\textwidth]{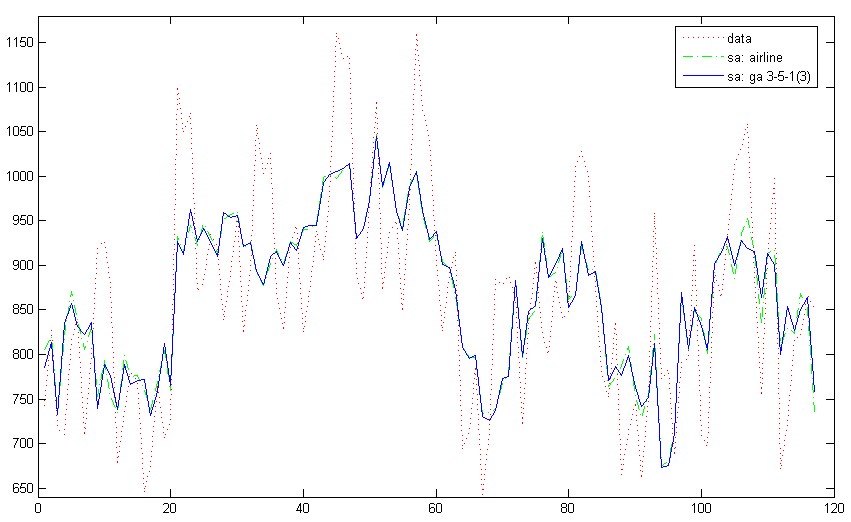}
\caption{Seasonal adjustment with airline and generalized airline models}
\label{fig:media}
\end{figure}

The same seasonal adjustment procedure is then done with the generalized airline model, using parameter estimates from the airline model as initial parameters:
\begin{small}\begin{verbatim}
param0 = air.param([1 2 2 3]); param0(1:3) = -param0(1:3); param0(2:3) = param0(2:3).^(1/12);
ga = estimate(y, ssmodel('genair', 3, 5, 3), param0);
gacom = ssmhtd(ga);
ycom = signal(statesmo(y, gacom), gacom);
gaseas = ycom(2, :);
\end{verbatim}\end{small}
The code creates a generalized airline 3-5-1 model, Hillmer-Tiao decomposition produces the same components as for the airline model since the two models have the same order. From the code it can be seen that the various functions work transparently across different ARIMA type models. Figure \ref{fig:media} shows the comparison between the two seasonal adjustment results.

\bibliographystyle{unsrt}
\bibliography{ssmpaper}

\appendix

\section{Predefined Model Reference}
\label{app:predefined}

The predefined models can be organized into two categories, observation disturbance models and normal models. The former contains only specification of the observation disturbance, and is used primarily for model combination; the latter contains all other models, and can in turn be partitioned into structural time series models, ARIMA type models, and other models.

The following is a list of each model, their string code and the arguments required to construct them via the constructor \texttt{ssmodel}.
\begin{itemize}
\item \textbf{Observation disturbance models}
    \begin{itemize}
      \item \textbf{Gaussian noise:} \texttt{$\sim$\footnote{Substitute with name of relevant the class constructor.}('Gaussian' | 'normal'[, p, cov])}\\
            \texttt{p} is the number of variables, default \texttt{1}.\\
            \texttt{cov} is true if they are correlated, default \texttt{true}.
      \item \textbf{Null noise:} \texttt{$\sim$('null'[, p])}\\
            \texttt{p} is the number of variables, default \texttt{1}.
      \item \textbf{Poisson error:} \texttt{$\sim$('poisson')}
      \item \textbf{Binary error:} \texttt{$\sim$('binary')}
      \item \textbf{Binomial error:} \texttt{$\sim$('binomial', k)}\\
            \texttt{k} is the number of trials, can be a scalar or row vector.
      \item \textbf{Negative binomial error:} \texttt{$\sim$('negbinomial', k)}\\
            \texttt{k} is the number of trials, can be a scalar or row vector.
      \item \textbf{Exponential error:} \texttt{$\sim$('exp')}
      \item \textbf{Multinomial error:} \texttt{$\sim$('multinomial', h, k)}\\
            \texttt{h} is the number of cells.\\
            \texttt{k} is the number of trials, can be a scalar or row vector.
      \item \textbf{Exponential family error:} \texttt{$\sim$('expfamily', b, d2b, id2bdb, c)}
            \[p(y|\theta)=\exp\left(y^T\theta-b(\theta)+c(y)\right)\]
            \texttt{b} is the function $b(\theta)$.\\
            \texttt{d2b} is $\ddot{b}(\theta)$, the second derivative of $b(\theta)$.\\
            \texttt{id2bdb} is $\ddot{b}(\theta)^{-1}\dot{b}(\theta)$.\\
            \texttt{c} is the function $c(y)$.
      \item \textbf{Zero-mean stochastic volatility error:} \texttt{$\sim$('zmsv')}
      \item \textbf{t-distribution noise:} \texttt{$\sim$('t'[, nu])}\\
            \texttt{nu} is the degree of freedom, will be estimated as model parameter if not specified.
      \item \textbf{Gaussian mixture noise:} \texttt{$\sim$('mix')}
      \item \textbf{General error noise:} \texttt{$\sim$('error')}
    \end{itemize}
\item \textbf{Structural time series models}
    \begin{itemize}
      \item \textbf{Integrated random walk:} \texttt{$\sim$('irw', d)}\\
            \texttt{d} is the order of integration.
      \item \textbf{Local polynomial trend:} \texttt{$\sim$('lpt', d)}\\
            \texttt{d} is the order of the polynomial.
      \item \textbf{Local level model:} \texttt{$\sim$('llm')}
      \item \textbf{Local level trend:} \texttt{$\sim$('llt')}
      \item \textbf{Seasonal components:} \texttt{$\sim$('seasonal', type, s)}\\
            \texttt{type} can be \texttt{'dummy'}, \texttt{'dummy fixed'}, \texttt{'h\&s'}, \texttt{'trig1'}, \texttt{'trig2'} or \texttt{'trig fixed'}.\\
            \texttt{s} is the seasonal period.
      \item \textbf{Cycle component:} \texttt{$\sim$('cycle')}
      \item \textbf{Regression components:} \texttt{$\sim$('reg', x[, varname])}\\
      \item \textbf{Dynamic regression components:} \texttt{$\sim$('dynreg', x[, varname])}\\
            \texttt{x} is a $m\times n$ matrix, $m$ is the number of regression variables.\\
            \texttt{varname} is the name of the variables.
      \item \textbf{Intervention components:} \texttt{$\sim$('intv', n, type, tau)}\\
            \texttt{n} is the total time duration.\\
            \texttt{type} can be \texttt{'step'}, \texttt{'pulse'}, \texttt{'slope'} or \texttt{'null'}.\\
            \texttt{tau} is the onset time.
      \item \textbf{Constant components:} \texttt{$\sim$('constant')}
      \item \textbf{Trading day variables:} \texttt{$\sim$('td6' | 'td1', y, m, N)}\\
            \texttt{'td6'} creates a six-variable trading day model.\\
            \texttt{'td1'} creates a one-variable trading day model.
      \item \textbf{Length-of-month variables:} \texttt{$\sim$('lom', y, m, N)}
      \item \textbf{Leap-year variables:} \texttt{$\sim$('ly', y, m, N)}
      \item \textbf{Easter effect variables:} \texttt{$\sim$('ee', y, m, N, d)}\\
            \texttt{y} is the starting year.\\
            \texttt{m} is the starting month.\\
            \texttt{N} is the total number of months.\\
            \texttt{d} is the number of days before Easter.
      \item \textbf{Structural time series models:} \texttt{$\sim$('stsm', lvl, seas, s[, cycle, x])}\\
            \texttt{lvl} is \texttt{'level'} or \texttt{'trend'}.\\
            \texttt{seas} is the seasonal type (see seasonal components).\\
            \texttt{s} is the seasonal period.\\
            \texttt{cycle} is \texttt{true} if there's a cycle component in the model, default \texttt{false}.\\
            \texttt{x} is explanatory (regression) variables (see regression components).
      \item \textbf{Common levels models:} \texttt{$\sim$('commonlvls', p, A\_ast, a\_ast)}
            \begin{eqnarray*}
            y_t & = & \left[\begin{array}{c}0 \\ a^\ast\end{array}\right] + \left[\begin{array}{c}I_r \\ A^\ast\end{array}\right]\mu_t^\ast+\varepsilon_t \\
            \mu_{t+1}^\ast & = & \mu_t^\ast + \eta_t^\ast
            \end{eqnarray*}
            \texttt{p} is the number of variables (length of $y_t$).\\
            \texttt{A\_ast} is $A^\ast$, a $(p-r)\times r$ matrix.\\
            \texttt{a\_ast} is $a^\ast$, a $(p-r)\times1$ vector.
      \item \textbf{Multivariate local level models:} \texttt{$\sim$('mvllm', p[, cov])}
      \item \textbf{Multivariate local level trend:} \texttt{$\sim$('mvllt', p[, cov])}
      \item \textbf{Multivariate seasonal components:} \texttt{$\sim$('mvseasonal', p, cov, type, s)}
      \item \textbf{Multivariate cycle component:} \texttt{$\sim$('mvcycle', p[, cov])}\\
            \texttt{p} is the number of variables.\\
            \texttt{cov} is a logical vector that is true for each correlated disturbances, default all \texttt{true}.\\
            Other arguments are the same as the univariate versions.
      \item \textbf{Multivariate regression components:} \texttt{$\sim$('mvreg', p, x[, dep])}\\
            \texttt{dep} is a $p\times m$ logical matrix that specifies dependence of data variables on regression variables.
      \item \textbf{Multivariate intervention components:} \texttt{$\sim$('mvintv', p, n, type, tau)}
      \item \textbf{Multivariate structural time series models:} \texttt{$\sim$('mvstsm', p, cov, lvl, seas, s[, cycle, x, dep])}\\
            \texttt{cov} is a logical vector that specifies the covariance structure of each component in turn.
    \end{itemize}
\item \textbf{ARIMA type models}
    \begin{itemize}
      \item \textbf{ARMA models:} \texttt{$\sim$('arma', p, q, mean)}
      \item \textbf{ARIMA models:} \texttt{$\sim$('arima', p, d, q, mean)}
      \item \textbf{Multiplicative seasonal ARIMA models:} \texttt{$\sim$('sarima', p, d, q, P, D, Q, s, mean)}
      \item \textbf{Seasonal sum ARMA models:} \texttt{$\sim$('sumarma', p, q, D, s, mean)}
      \item \textbf{SARIMA with Hillmer-Tiao decomposition:} \texttt{$\sim$('sarimahtd', p, d, q, P, D, Q, s, gauss)}\\
            \texttt{p} is the order of AR.\\
            \texttt{d} is the order of I.\\
            \texttt{q} is the order of MA.\\
            \texttt{P} is the order of seasonal AR.\\
            \texttt{D} is the order of seasonal I.\\
            \texttt{Q} is the order of seasonal MA.\\
            \texttt{s} is the seasonal period.\\
            \texttt{mean} is true if the model has mean.\\
            \texttt{gauss} is true if the irregular component is Gaussian.
      \item \textbf{Airline models:} \texttt{$\sim$('airline'[, s])}\\
            \texttt{s} is the period, default 12.
      \item \textbf{Generalized airline models:} \texttt{$\sim$('genair', nparam, nfreq, freq)}\\
            \texttt{nparam} is the number of parameters, \texttt{3} or \texttt{4}.\\
            \texttt{nfreq} is the size of the largest subset of frequencies sharing the same parameter.\\
            \texttt{freq} is an array containing the members of the smallest subset.
      \item \textbf{ARIMA component models:} \texttt{$\sim$('arimacom', d, D, s, phi, theta, ksivar)}\\
            The arguments match those of the function \texttt{htd}, see its description for details.
    \end{itemize}
\item \textbf{Other models}
    \begin{itemize}
      \item \textbf{Cubic spline smoothing (continuous time):} \texttt{$\sim$('spline', delta)}\\
            \texttt{delta} is the time duration of each data point.
      \item \textbf{1/f noise models (approximated by AR):} \texttt{$\sim$('1/f noise', m)}\\
            \texttt{m} is the order of the approximating AR process.
    \end{itemize}
\end{itemize}

\section{Class Reference}
\label{app:class}

\subsection{SSMAT}

The class \texttt{SSMAT} represents state space matrices, can be stationary or dynamic. \texttt{SSMAT} can be horizontally, vertically, and block diagonally concatenated with each other, as well as with ordinary matrices. Parenthesis reference can also be used, where two or less indices indicate the stationary part, and three indices indicates a 3-d matrix with time as the third dimension.

\subsubsection{Subscripted reference and assignment}

\begin{tabular}{llll}
\tt .mat      & RA\footnotemark                            & Stationary part of \texttt{SSMAT}     & Assignment cannot change size.\\
\tt .mmask    & R  & Variable mask                         & A logical matrix the same size as \texttt{mat}.\\
\tt .n        & R  & Time duration                         & Is equal to \texttt{size(dvec, 2)}.\\
\tt .dmmask   & R  & Dynamic mask                          & A logical matrix the same size as \texttt{mat}.\\
\tt .d        & R  & Number of dynamic elements            & Is equal to \texttt{size(dvec, 1)}.\\
\tt .dvec     & RA & Dynamic vector sequence               & A \texttt{d}${}\times{}$\texttt{n} matrix. Assignment cannot change \texttt{d}.\\
\tt .dvmask   & R  & Dynamic vector variable mask          & A \texttt{d}${}\times1$ logical vector.\\
\tt .const    & R  & True if \texttt{SSMAT} is constant    & A logical scalar.\\
\tt .sta      & R  & True if \texttt{SSMAT} is stationary  & A logical scalar.\\
\tt (i, j)    & RA & Stationary part of \texttt{SSMAT}     & Assignment cannot change size. \\
\tt (i, j, t) & RA & \texttt{SSMAT} as a 3-d matrix        & Assignment cannot change size.
\end{tabular}
\footnotetext{R - reference, A - assignment.}

\subsubsection{Class methods}

\begin{itemize}
\item \texttt{\textbf{ssmat}}\\
      \texttt{SSMAT} constructor.\\
      \texttt{SSMAT(m[, mmask])} creates a \texttt{SSMAT}. \texttt{m} can be a 2-d or 3-d matrix, where the third dimension is time. \texttt{mmask} is a 2-d logical matrix masking the variable part of matrix, which can be omitted or set to \texttt{[]} for constant \texttt{SSMAT}.\\
      \texttt{SSMAT(m, mmask, dmmask[, dvec, dvmask])} creates a dynamic \texttt{SSMAT}. \texttt{m} is a 2-d matrix forming the stationary part of the \texttt{SSMAT}. \texttt{mmask} is a logical matrix masking the variable part of matrix, which can be set to \texttt{[]} for constant \texttt{SSMAT}. \texttt{dmmask} is a logical matrix masking the dynamic part of \texttt{SSMAT}. \texttt{dvec} is a \texttt{nnz(dmmask)}${}\times{}$\texttt{n} matrix, \texttt{dvec(:, t)} is the values for the dynamic part at time \texttt{t} (\texttt{m(dmmask) = dvec(:, t)} for time \texttt{t}), default is a \texttt{nnz(dmmask)}${}\times1$ zero matrix. \texttt{dvmask} is a \texttt{nnz(dmmask)}${}\times1$ logical vector masking the variable part of \texttt{dvec}.
\item \texttt{\textbf{isconst}}\\
      \texttt{ISCONST($\cdot$\footnote{An instance of the class in question.})} returns true if the stationary part of \texttt{SSMAT} is constant.
\item \texttt{\textbf{isdconst}}\\
      \texttt{ISDCONST($\cdot$)} returns true if the dynamic part of \texttt{SSMAT} is constant.
\item \texttt{\textbf{issta}}\\
      \texttt{ISSTA($\cdot$)} returns true if \texttt{SSMAT} is stationary.
\item \texttt{\textbf{size}}\\
      \texttt{[ \ldots\ ] = SIZE($\cdot$, \ldots)} is equivalent to \texttt{[ \ldots\ ] = SIZE($\cdot$.mat, \ldots)}.
\item \texttt{\textbf{getmat}}\\
      \texttt{GETMAT($\cdot$)} returns \texttt{SSMAT} as a matrix if stationary, or as a cell array of matrices if dynamic.
\item \texttt{\textbf{getdvec}}\\
      \texttt{GETDVEC($\cdot$, mmask)} returns the dynamic elements specified in matrix mask \texttt{mmask} as a vector sequence.
\item \texttt{\textbf{getn}}\\
      \texttt{GETN($\cdot$)} returns the time duration of \texttt{SSMAT}.
\item \texttt{\textbf{setmat}}\\
      \texttt{SETMAT($\cdot$, vec)} updates the variable stationary matrix, \texttt{vec} must have size \texttt{nnz(mmask)}${}\times1$.
\item \texttt{\textbf{setdvec}}\\
      \texttt{SETDVEC($\cdot$, dsubvec)} updates the variable dynamic vector sequence, \texttt{dsubvec} must have \texttt{nnz(dvmask)} rows.
\item \texttt{\textbf{plus}}\\
      \texttt{SSMAT} addition.
\end{itemize}

\subsection{SSDIST}

The class \texttt{SSDIST} represents state space distributions, a set of non-Gaussian distributions that governs a disturbance vector (which can also have Gaussian elements), and is a child class of \texttt{SSMAT}. \texttt{SSDIST} are constrained to be square matrices, hence can only be block diagonally concatenated, any other types of concatenation will ignore the non-Gaussian distribution part. Predefined non-Gaussian distributions can be constructed via the \texttt{SSDIST} constructor.

\subsubsection{Subscripted reference and assignment}

\begin{tabular}{llll}
\tt .nd       & R  & Number of distributions                         & A scalar.\\
\tt .type     & R  & Type of each distribution                       & A 0-1 row vector.\\
\tt .matf     & R  & Approximation functions    & A cell array of functions.\\
\tt .logpf    & R  & Log probability functions   & A cell array of functions.\\
\tt .diagmask & R  & Non-Gaussian masks                  & A \texttt{size(mat, 1)}${}\times{}$\texttt{nd} logical matrix. \\
\tt .dmask    & R  & Variable mask for the distributions             & A $1\times{}$\texttt{nd} logical vector.\\
\tt .const    & R  & True if \texttt{SSDIST} is constant             & A logical scalar.
\end{tabular}

\subsubsection{Class methods}

\begin{itemize}
\item \texttt{\textbf{ssdist}}\\
      \texttt{SSDIST} constructor.\\
      \texttt{SSDIST('', type[, matf, logpf])} creates a single univariate non-Gaussian distribution. \texttt{type} is 0 for exponential family distributions and 1 for additive noise distributions. \texttt{matf} is the function that generates the approximated Gaussian variance matrix given observations and signal or disturbances. \texttt{logpf} is the function that calculates the log probability of observations given observation and signal or disturbances. If the last two arguments are omitted, then the \texttt{SSDIST} is assumed to be variable. \texttt{SSDIST} with multiple non-Gaussian distributions can be formed by constructing a \texttt{SSDIST} for each and then block diagonally concatenating them.
\item \texttt{\textbf{isdistconst}}\\
      \texttt{ISDISTCONST($\cdot$)} returns true if the non-Gaussian part of \texttt{SSDIST} is constant.
\item \texttt{\textbf{setdist}}\\
      \texttt{SETDIST($\cdot$, distf)} updates the non-Gaussian distributions. \texttt{distf} is a cell matrix of functions.
\item \texttt{\textbf{setgauss}}\\
      \texttt{SETGAUSS($\cdot$, eps)} or\\
      \texttt{[$\cdot$ ytilde] = SETGAUSS($\cdot$, y, theta)} calculates the approximating Gaussian covariance, the first form is used if there are no exponential family distributions, and the data \texttt{y} need to be transformed into \texttt{ytilde} in the second form if there are.
\item \texttt{\textbf{logprobrat}}\\
      \texttt{LOGPROBRAT($\cdot$, N, eps)} or\\
      \texttt{LOGPROBRAT($\cdot$, N, y, theta, eps)} calculates the log probability ratio between the original non-Gaussian distribution and the approximating Gaussian distribution of the observations, it is used when calculating the non-Gaussian loglikelihood by importance sampling. The first form is used if there are no exponential family distributions. \texttt{N} is the number of importance samples.
\end{itemize}

\subsubsection{Predefined non-Gaussian distributions}

\begin{itemize}
\item \textbf{Poisson distribution:} \texttt{$\sim$('poisson')}
\item \textbf{Binary distribution:} \texttt{$\sim$('binary')}
\item \textbf{Binomial distribution:} \texttt{$\sim$('binomial', k)}
\item \textbf{Negative binomial distribution:} \texttt{$\sim$('negbinomial', k)}\\
      \texttt{k} is the number of trials, can be a scalar or row vector.
\item \textbf{Exponential distribution:} \texttt{$\sim$('exp')}
\item \textbf{Multinomial distribution:} \texttt{$\sim$('multinomial', h, k)}\\
      \texttt{h} is the number of cells.\\
      \texttt{k} is the number of trials, can be a scalar or row vector.
\item \textbf{General exponential family distribution:} \texttt{$\sim$('expfamily', b, d2b, id2bdb, c)}\\
      \[p(y|\theta)=\exp\left(y^T\theta-b(\theta)+c(y)\right)\]
      \texttt{b} is the function $b(\theta)$.\\
      \texttt{d2b} is $\ddot{b}(\theta)$, the second derivative of $b(\theta)$.\\
      \texttt{id2bdb} is $\ddot{b}(\theta)^{-1}\dot{b}(\theta)$.\\
      \texttt{c} is the function $c(y)$.
\end{itemize}

\subsection{SSFUNC}

The class \texttt{SSFUNC} represents state space functions, a set of nonlinear functions that describe state vector transformations, and is a child class of \texttt{SSMAT}. Since linear transformations in SSM are represented by left multiplication of matrices, it is only possible for nonlinear functions (and hence it's matrix approximation) to be either block diagonally or horizontally concatenated. The former concatenation method produces multiple resulting output vectors, whereas the latter produces the sum of the output vectors of all functions as output. Vertical concatenation will ignore the nonlinear part of \texttt{SSFUNC}.

\subsubsection{Subscripted reference and assignment}

\begin{tabular}{llll}
\tt .nf       & R  & Number of functions         & A scalar.\\
\tt .f        & R  & The nonlinear functions     & A cell array of functions.\\
\tt .df       & R  & Derivative of each function & A cell array of functions.\\
\tt .horzmask & R  & Nonlinear function output masks  & A \texttt{size(mat, 1)}${}\times{}$\texttt{nf} logical matrix. \\
\tt .vertmask & R  & Nonlinear function input masks   & A \texttt{size(mat, 2)}${}\times{}$\texttt{nf} logical matrix. \\
\tt .fmask    & R  & Variable mask for the functions      & A $1\times{}$\texttt{nf} logical vector.\\
\tt .const    & R  & True if \texttt{SSFUNC} is constant     & A logical scalar.
\end{tabular}

\subsubsection{Class methods}

\begin{itemize}
\item \texttt{\textbf{ssfunc}}\\
      \texttt{SSFUNC} constructor.\\
      \texttt{SSFUNC('', p, m[, f, df])} creates a single nonlinear function. \texttt{p} is the input vector dimension and \texttt{m} is the output vector dimension. Hence the approximating linear matrix will be \texttt{p}${}\times{}$\texttt{m}. \texttt{f} and \texttt{df} is the nonlinear function and its derivative. \texttt{f} maps \texttt{m}${}\times1$ vectors and time \texttt{t} to \texttt{p}${}\times1$ vectors, and \texttt{df} maps \texttt{m}${}\times1$ vectors and time \texttt{t} to \texttt{p}${}\times{}$\texttt{m} matrices. If \texttt{f} and \texttt{df} are not both provided, the \texttt{SSFUNC} is assumed to be variable. \texttt{SSFUNC} with multiple functions can be constructed by defining individual \texttt{SSFUNC} for each function, and concatenating them.
\item \texttt{\textbf{isfconst}}\\
      \texttt{ISFCONST($\cdot$)} returns true if the nonlinear part of \texttt{SSFUNC} is constant.
\item \texttt{\textbf{getfunc}}\\
      \texttt{GETFUNC($\cdot$, x, t)} returns \texttt{f(x, t)}, the transform of \texttt{x} at time \texttt{t}.
\item \texttt{\textbf{getvertmask}}\\
      \texttt{GETVERTMASK($\cdot$)} returns \texttt{vertmask}.
\item \texttt{\textbf{setfunc}}\\
      \texttt{SETFUNC($\cdot$, funcf)} updates the nonlinear functions. \texttt{funcf} is a cell matrix of functions.
\item \texttt{\textbf{setlinear}}\\
      \texttt{[$\cdot$ c] = SETLINEAR($\cdot$, alpha)} calculates the approximating linear matrix, \texttt{c} is an additive constant in the approximation.
\end{itemize}

\subsection{SSPARAM}

The class \texttt{SSPARAM} represents state space model parameters. \texttt{SSPARAM} can be seen as a row vector of parameter values, hence they are combined by horizontal concatenation with a slight difference, as described below.

\subsubsection{Subscripted reference and assignment}

\begin{tabular}{llll}
\tt .w     & R  & Number of parameters         & A scalar.\\
\tt .name  & R  & Parameter names              & A cell array of strings.\\
\tt .value & RA & Transformed parameter values & A $1\times{}$\texttt{w} vector. Assignment cannot change size.
\end{tabular}

\subsubsection{Class methods}

\begin{itemize}
\item \texttt{\textbf{ssparam}}\\
      \texttt{SSPARAM} constructor.\\
      \texttt{SSPARAM(w[, transform])} creates a \texttt{SSPARAM} with \texttt{w} parameters.\\
      \texttt{SSPARAM(name[, transform])} creates a \texttt{SSPARAM} with parameter names \texttt{name}, which is a cell array of strings.\\
      \texttt{transform} is a cell array of strings describing transforms used for each parameter.\\
      \texttt{SSPARAM(name, group, transform)} creates a \texttt{SSPARAM} with parameter names specified by \texttt{name}. \texttt{group} is a row vector that specifies the number of parameters included in each parameter subgroup, and \texttt{transform} is of the same length which specifies transforms for each group.
\item \texttt{\textbf{get}}\\
      \texttt{GET($\cdot$)} returns the untransformed parameter values.
\item \texttt{\textbf{set}}\\
      \texttt{SET($\cdot$, value)} sets the untransformed parameter values.
\item \texttt{\textbf{remove}}\\
      \texttt{REMOVE($\cdot$, mask)} removes the parameters specified by \texttt{mask}.
\item \texttt{\textbf{horzcat}}\\
      \texttt{[$\cdot$ pmask] = HORZCAT($\cdot_1$, $\cdot_2$, \ldots)} combines multiple \texttt{SSPARAM} into one, and optionally generates a cell array of parameter masks \texttt{pmask}, masking each original parameter sets with respect to the new combined parameter set.
\end{itemize}

\subsection{SSMODEL}

The class \texttt{SSMODEL} represents state space models and embeds all the previous classes. \texttt{SSMODEL} can be combined by horizontal or block diagonal concatenation. The former is the basis for additive model combination, where the observation is the sum of individual signals. The latter is an independent combination in which the observation is the vertical concatenation of the individual signals, which obviously needs to be further combined or changed to be useful. Only \texttt{SSMODEL} objects are used directly in data analysis functions, knowledge of embedded objects are needed only when defining custom models. For a list of predefined models see appendix \ref{app:predefined}.

\subsubsection{Subscripted reference and assignment}

\begin{tabular}{llll}
\tt .name  & RA & Model name                        & A string.\\
\tt .Hinfo & R  & Observation disturbance info      & A structure.\\
\tt .info  & R  & Model component info              & A cell array of structures.\\
\tt .p     & R  & Model observation dimension       & A scalar.\\
\tt .m     & R  & Model state dimension             & A scalar.\\
\tt .r     & R  & Model state disturbance dimension & A scalar.\\
\tt .n     & R  & Model time duration               & A scalar. Value is \texttt{1} for stationary models.\\
\tt .H     & R  & Observation disturbance           & A \texttt{SSMAT} or \texttt{SSDIST}.\\
\tt .Z     & R  & State to observation transform    & A \texttt{SSMAT} or \texttt{SSFUNC}.\\
\tt .T     & R  & State update transform            & A \texttt{SSMAT} or \texttt{SSFUNC}.\\
\tt .R     & R  & Disturbance to state transform    & A \texttt{SSMAT}.\\
\tt .Q     & R  & State disturbance                 & A \texttt{SSMAT} or \texttt{SSDIST}.\\
\tt .c     & R  & State update constant             & A \texttt{SSMAT}.\\
\tt .a1    & RA & Initial state mean                & A stationary \texttt{SSMAT}, can assign matrices.\\
\tt .P1    & RA & Initial state variance            & A stationary \texttt{SSMAT}, can assign matrices.\\
\tt .sta   & R  & True for stationary \texttt{SSMODEL} & A logical scalar.\\
\tt .linear& R  & True for linear \texttt{SSMODEL}  & A logical scalar.\\
\tt .gauss & R  & True for Gaussian \texttt{SSMODEL}& A logical scalar.\\
\tt .w     & R  & Number of parameters              & A scalar.\\
\tt .paramname & R & Parameter names                & A cell array of strings.\\
\tt .param & RA & Parameter values                  & A $1\times{}$\texttt{w} vector. Assignment cannot change \texttt{w}.\\
\tt .psi   & RA & Transformed parameter values      & A $1\times{}$\texttt{w} vector. Assignment cannot change \texttt{w}.\\
\tt .q     & R  & Number of initial diffuse elements & A scalar.
\end{tabular}

\subsubsection{Class Methods}

\begin{itemize}
\item \texttt{\textbf{ssmodel}}\\
      \texttt{SSMODEL} constructor.\\
      \texttt{SSMODEL('', info, H, Z, T, R, Q)} creates a constant state space model with complete diffuse initialization and \texttt{c}${}=0$.
      \texttt{SSMODEL('', info, H, Z, T, R, Q, A, func, grad, psi[, pmask, P1, a1, c])} create a state space model. \texttt{info} is a string or a free form structure with field 'type' describing the model component. \texttt{H} is a \texttt{SSMAT} or \texttt{SSDIST}. \texttt{Z} is a \texttt{SSMAT} or \texttt{SSFUNC}. \texttt{T} is a \texttt{SSMAT} or \texttt{SSFUNC}. \texttt{R} is a \texttt{SSMAT}. \texttt{Q} is a \texttt{SSMAT} or \texttt{SSDIST}. \texttt{A} is a cell array of strings or a single string, each corresponding to \texttt{func}, a cell array of functions or a single function. Each string of \texttt{A} is a concatenation of some of \texttt{'H'}, \texttt{'Z'}, \texttt{'T'}, \texttt{'R'}, \texttt{'Q'}, \texttt{'c'}, \texttt{'a1'}, \texttt{'P1'}, \texttt{'Hd'}, \texttt{'Zd'}, \texttt{'Td'}, \texttt{'Rd'}, \texttt{'Qd'}, \texttt{'cd'}, \texttt{'Hng'}, \texttt{'Qng'}, \texttt{'Znl'}, \texttt{'Tnl'}, representing each part of each element of \texttt{SSMODEL}. For example, \texttt{'Qng'} specifies the non-Gaussian part of \texttt{Q}, and \texttt{'Zd'} specifies the dynamic part of \texttt{Z}. A complicated example of a string in \texttt{A} is
\begin{center}
\texttt{['H' repmat('ng', 1, ~gauss) repmat('T', 1, p+P>0) 'RQP1']},
\end{center}
which is the adjacency string for a possibly non-Gaussian ARIMA components model.
\texttt{func} is a cell array of functions that updates model matrices. If multiple parts of the model are updated the output for each part must be ordered as for the adjacency matrix, but parts not updated can be skipped, for example, a function that updates \texttt{H}, \texttt{T} and \texttt{Q} must output \texttt{[Hvec Tvec Qvec]} in this order. \texttt{grad} is the derivative of \texttt{func}, each of which can be \texttt{[]} if not differentiable. For the example function above the gradient function would output \texttt{[Hvec Tvec Qvec Hgrad Tgrad Qgrad]} in order. \texttt{psi} is a \texttt{SSPARAM}. \texttt{pmask} is a cell array of parameter masks for the corresponding functions, can be omitted or set to [] if there's only one update function (which presumably uses all parameters). \texttt{P1} and \texttt{a1} are both stationary \texttt{SSMAT}. \texttt{c} is a \texttt{SSMAT}. All arguments expecting \texttt{SSMAT} can also take numeric matrices.
\item \texttt{\textbf{isgauss}}\\
      \texttt{ISGAUSS($\cdot$)} is true for Gaussian \texttt{SSMODEL}.
\item \texttt{\textbf{islinear}}\\
      \texttt{ISLINEAR($\cdot$)} is true for linear \texttt{SSMODEL}.
\item \texttt{\textbf{issta}}\\
      \texttt{ISSTA($\cdot$)} is true for stationary \texttt{SSMODEL}. Note that all \texttt{SSFUNC} are implicitly assumed to be dynamic, but are treated as stationary by this and similar functions.
\item \texttt{\textbf{set}}\\
      \texttt{SET($\cdot$, A, M, func, grad, psi, pmask)} changes one specific element of \texttt{SSMODEL}. \texttt{A} is an adjacency string, specifying which elements the function or functions \texttt{func} updates. \texttt{A} is a adjacency string as described earlier, corresponding to \texttt{func}. \texttt{M} should be the new value for one of \texttt{H}, \texttt{Z}, \texttt{T}, \texttt{R}, \texttt{Q}, \texttt{c}, \texttt{a1} or \texttt{P1}, and \texttt{A} is constrained to contain only references to that element. \texttt{grad} is the same type as \texttt{func} and is its gradient, set individual functions to 0 if gradient does not exist or needed. \texttt{psi} is a \texttt{SSPARAM} that stores all the parameters used by functions in \texttt{func}, and \texttt{pmask} is the parameter mask for each function.
\item \texttt{\textbf{setparam}}\\
      \texttt{SETPARAM($\cdot$, psi, transformed)} sets the parameters for \texttt{SSMODEL}. If \texttt{transformed} is true, \texttt{psi} is the new value for the transformed parameters, else \texttt{psi} is the new value for the untransformed parameters.
\end{itemize}

\section{Function Reference}

\subsection{User-defined update functions}

This section details the format of update functions for various parts of classes \texttt{SSMAT}, \texttt{SSDIST} and \texttt{SSFUNC}.

\begin{itemize}
\item \textbf{Stationary part}\\
      \texttt{vec = func(psi)} updates the stationary part of \texttt{SSMAT}. \texttt{vec} must be a column vector such that \texttt{mat(mmask) = vec} would correctly update the stationary matrix \texttt{mat} given parameters \texttt{psi}.
\item \textbf{Dynamic part}\\
      \texttt{dsubvec = func(psi)} updates the dynamic part of \texttt{SSMAT}. \texttt{dsubvec} must be a matrix such that \texttt{dvec(dvmask, 1:size(dsubvec, 2)) = dsubvec} would correctly update the dynamic vector sequence \texttt{dvec} given parameters \texttt{psi}.
\item \textbf{Non-Gaussian part}\\
      \texttt{distf = func(psi)} updates the non-Gaussian part of \texttt{SSDIST}. \texttt{distf} must be a cell matrix of function handles such that \texttt{[matf{dmask}] = distf{:, 1}} and \texttt{[logpf{dmask}] = distf{:, 2}} would correctly update the distributions given parameters \texttt{psi}.
\item \textbf{Nonlinear part}\\
      \texttt{funcf = func(psi)} updates the nonlinear part of \texttt{SSFUNC}. \texttt{funcf} must be a cell matrix of function handles such that \texttt{[f{fmask}] = funcf{:, 1}} and \texttt{[df{fmask}] = funcf{:, 2}} would correctly update the nonlinear functions given parameters \texttt{psi}.
\end{itemize}

Note that an update function can return more than one of the four kind of output, thus updating multiple part of a \texttt{SSMODEL}, but the order of the output arguments must follow the convention: stationary part of \texttt{H}, \texttt{Z}, \texttt{T}, \texttt{R}, \texttt{Q}, \texttt{c}, \texttt{a1}, \texttt{P1}, dynamic part of \texttt{H}, \texttt{Z}, \texttt{T}, \texttt{R}, \texttt{Q}, \texttt{c}, non-Gaussian part of \texttt{H}, \texttt{Q}, nonlinear part of \texttt{Z}, \texttt{T}, with possible omissions.

\subsection{Data analysis functions}

Most functions in this section accepts analysis settings options, specified as option name and option value pairs (e.g. \texttt{('disp', 'off')}). These groups of arguments are specified at the end of each function that accepts them, and are represented by \texttt{opt} in this section.

\begin{itemize}
\item \texttt{\textbf{batchsmo}}\\
      \texttt{[alphahat epshat etahat] = BATCHSMO(y, model[, opt])} performs batch smoothing of multiple data sets. \texttt{y} is the data of dimension \texttt{p}${}\times{}$\texttt{n}${}\times{}$\texttt{N}, where \texttt{n} is the data length and \texttt{N} is the number of data sets, there must be no missing values. \texttt{model} is a \texttt{SSMODEL}. The output is respectively the smoothed state, smoothed observation disturbance and smoothed state disturbance, each of dimensions \texttt{m}${}\times{}$\texttt{n}${}\times{}$\texttt{N}, \texttt{p}${}\times{}$\texttt{n}${}\times{}$\texttt{N} and \texttt{r}${}\times{}$\texttt{n}${}\times{}$\texttt{N}. This is equivalent to doing \texttt{fastsmo} on each data set.
\item \texttt{\textbf{disturbsmo}}\\
      \texttt{[epshat etahat epsvarhat etavarhat] = DISTURBSMO(y, model[, opt])} performs disturbance smoothing. \texttt{y} is the data of dimension \texttt{p}${}\times{}$\texttt{n}, and \texttt{model} is a \texttt{SSMODEL}. The output is respectively the smoothed observation disturbance (\texttt{p}${}\times{}$\texttt{n}), smoothed state disturbance (\texttt{r}${}\times{}$\texttt{n}), smoothed observation disturbance variance (\texttt{p}${}\times{}$\texttt{p}${}\times{}$\texttt{n} or $1\times{}$\texttt{n} if \texttt{p}${}=1$) and smoothed state disturbance variance (\texttt{r}${}\times{}$\texttt{r}${}\times{}$\texttt{n} or $1\times{}$\texttt{n} if \texttt{r}${}=1$).
\item \texttt{\textbf{estimate}}\\
      \texttt{[model logL output] = ESTIMATE(y, model[, param0, alpha0, opt])} estimates the parameters of \texttt{model} starting from the initial parameter value \texttt{param0}. \texttt{y} is the data of dimension \texttt{p}${}\times{}$\texttt{n}, and \texttt{model} is a \texttt{SSMODEL}. \texttt{param0} can be empty if the current parameter values of \texttt{model} is used as initial value, and a scalar \texttt{param0} sets all parameters to the same value. Alternatively \texttt{param0} can be a logical row vector specifying which parameters to estimate, or a \texttt{2}${}\times{}$\texttt{w} matrix with the first row as initial value and second row as estimated parameter mask. The initial state sequence estimate \texttt{alpha0} is needed only when \texttt{model} is non-Gaussian or nonlinear. \texttt{output} is a structure that contains optimization routine information, approximated observation sequence $\tilde{y}$ if non-Gaussian or nonlinear, and the AIC and BIC of the output \texttt{model}.
\item \texttt{\textbf{fastsmo}}\\
      \texttt{[alphahat epshat etahat] = fastsmo(y, model[, opt])} performs fast smoothing. \texttt{y} is the data of dimension \texttt{p}${}\times{}$\texttt{n}, and \texttt{model} is a \texttt{SSMODEL}. The output is respectively the smoothed state (\texttt{m}${}\times{}$\texttt{n}), smoothed observation disturbance (\texttt{p}${}\times{}$\texttt{n}), and smoothed state disturbance (\texttt{r}${}\times{}$\texttt{n}).
\item \texttt{\textbf{gauss}}\\
      \texttt{[model ytilde] = GAUSS(y, model[, alpha0, opt])} calculates the Gaussian approximation. \texttt{y} is the data of dimension \texttt{p}${}\times{}$\texttt{n}, and \texttt{model} is a \texttt{SSMODEL}. \texttt{alpha0} is the initial state sequence estimate and can be empty or omitted.
\item \texttt{\textbf{kalman}}\\
      \texttt{[a P v F] = KALMAN(y, model[, opt])} performs Kalman filtering. \texttt{y} is the data of dimension \texttt{p}${}\times{}$\texttt{n}, and \texttt{model} is a \texttt{SSMODEL}. The output is respectively the filtered state (\texttt{m}${}\times{}$\texttt{n+1}),  filtered state variance (\texttt{m}${}\times{}$\texttt{m}${}\times{}$\texttt{n+1}, or $1\times{}$\texttt{n+1} if \texttt{m}${}=1$), one-step prediction error (innovation) (\texttt{p}${}\times{}$\texttt{n}), one-step prediction variance (\texttt{p}${}\times{}$\texttt{p}${}\times{}$\texttt{n}, or $1\times{}$\texttt{n} if \texttt{p}${}=1$).
\item \texttt{\textbf{linear}}\\
      \texttt{[model ytilde] = LINEAR(y, model[, alpha0, opt])} calculates the linear approximation. \texttt{y} is the data of dimension \texttt{p}${}\times{}$\texttt{n}, and \texttt{model} is a \texttt{SSMODEL}. \texttt{alpha0} is the initial state sequence estimate and can be empty or omitted.
\item \texttt{\textbf{loglik}}\\
      \texttt{LOGLIK(y, model[, ytilde, opt])} returns the log likelihood of \texttt{model} given \texttt{y}. \texttt{y} is the data of dimension \texttt{p}${}\times{}$\texttt{n}, and \texttt{model} is a \texttt{SSMODEL}. \texttt{ytilde} is the approximating observation $\tilde{y}$ and is needed for some non-Gaussian or nonlinear models.
\item \texttt{\textbf{sample}}\\
      \texttt{[y alpha eps eta] = SAMPLE(model, n[, N])} generates observation samples from \texttt{model}. \texttt{model} is a \texttt{SSMODEL}, \texttt{n} specifies the sampling data length, and \texttt{N} specifies how many sets of data to generate. \texttt{y} is the sampled data of dimension \texttt{p}${}\times{}$\texttt{n}${}\times{}$\texttt{N}, \texttt{alpha}, \texttt{eps}, \texttt{eta} are respectively the corresponding sampled state (\texttt{m}${}\times{}$\texttt{n}${}\times{}$\texttt{N}), observation disturbance (\texttt{p}${}\times{}$\texttt{n}${}\times{}$\texttt{N}), and state disturbance (\texttt{r}${}\times{}$\texttt{n}${}\times{}$\texttt{N}).
\item \texttt{\textbf{signal}}\\
      \texttt{SIGNAL(alpha, model)} generates the signal for each component according to the state sequence and model specification. \texttt{alpha} is the state of dimension \texttt{m}${}\times{}$\texttt{n}, and \texttt{model} is a \texttt{SSMODEL}. The output is a cell array of data each with dimension \texttt{p}${}\times{}$\texttt{n}, or a \texttt{M}${}\times{}$\texttt{n} matrix where \texttt{M} is the number of components if \texttt{p}${}=1$.
\item \texttt{\textbf{simsmo}}\\
      \texttt{[alphatilde epstilde etatilde] = SIMSMO(y, model, N[, antithetic, opt])} generates observation samples from \texttt{model} conditional on data \texttt{y}. \texttt{y} is the data of dimension \texttt{p}${}\times{}$\texttt{n}, and \texttt{model} is a \texttt{SSMODEL}. \texttt{antithetic} should be set to \texttt{1} if antithetic variables are used. The output is respectively the sampled state sequence (\texttt{m}${}\times{}$\texttt{n}${}\times{}$\texttt{N}), sampled observation disturbance (\texttt{p}${}\times{}$\texttt{n}${}\times{}$\texttt{N}), and sampled state disturbance (\texttt{r}${}\times{}$\texttt{n}${}\times{}$\texttt{N}).
\item \texttt{\textbf{statesmo}}\\
      \texttt{[alphahat V] = STATESMO(y, model[, opt])} performs state smoothing. \texttt{y} is the data of dimension \texttt{p}${}\times{}$\texttt{n}, and \texttt{model} is a \texttt{SSMODEL}. The output is respectively the smoothed state (\texttt{m}${}\times{}$\texttt{n}), smoothed state variance (\texttt{m}${}\times{}$\texttt{m}${}\times{}$\texttt{n}, or $1\times{}$\texttt{n} if \texttt{p}${}=1$). If only the first output argument is specified, fast state smoothing is automatically performed instead.
\item \texttt{\textbf{arimaselect}}\\
      \texttt{[p d q mean] = ARIMASELECT(y)} or\\
      \texttt{[p d q P D Q mean] = ARIMASELECT(y, s)} performs TRAMO model selection. \texttt{y} is the data of dimension \texttt{p}${}\times{}$\texttt{n}, and \texttt{s} is the seasonal period. The first form selects among ARIMA models with or without a mean. The second form selects among SARIMA models with or without a mean.
\item \texttt{\textbf{armadegree}}\\
      \texttt{[p q] = ARMADEGREE(y[, mean, mr])} or\\
      \texttt{[p q P Q] = ARMADEGREE(y, s[, mean, mr, ms])} determines the degrees of ARMA or SARMA from observation \texttt{y}. \texttt{y} is the data of dimension \texttt{p}${}\times{}$\texttt{n}, and \texttt{s} is the seasonal period. \texttt{mean} is true if models with mean are used for the selection. \texttt{mr} and \texttt{ms} is the largest degree to consider for regular ARMA and seasonal ARMA respectively.
\item \texttt{\textbf{diffdegree}}\\
      \texttt{[d mean] = DIFFDEGREE(y[, ub, tsig])} or\\
      \texttt{[d D mean] = DIFFDEGREE(y, s[, ub, tsig])} performs unit root detection. The first form only considers regular differencing. \texttt{y} is the data of dimension \texttt{p}${}\times{}$\texttt{n}, \texttt{s} is the seasonal period, \texttt{ub} is the threshold for unit roots, default is \texttt{[0.97 0.88]}, and \texttt{tsig} is the t-value threshold for mean detection, default is \texttt{1.5}.
\item \texttt{\textbf{htd}}\\
      \texttt{[theta ksivar] = HTD(d, D, s, Phi, Theta[, etavar])} performs Hillmer-Tiao decomposition. \texttt{d} is the order of regular differencing, \texttt{D} is the order of seasonal differencing, \texttt{s} is the seasonal period, \texttt{Phi} is a cell array of increasing order polynomials representing autoregressive factors for each component, \texttt{Theta} is an increasing order polynomial representing the moving average factor, \texttt{etavar} is the disturbance variance. The output \texttt{theta} is a cell array of decomposed moving average factors for each component, and \texttt{ksivar} is the corresponding disturbance variance for each component (including the irregular variance at the end).
\item \texttt{\textbf{ssmhtd}}\\
      \texttt{SSMHTD(model)} performs Hillmer-Tiao decomposition on \texttt{SSMODEL} \texttt{model} and returns an ARIMA component model. This is a wrapper for the function \texttt{htd}.
\item \texttt{\textbf{loglevel}}\\
      \texttt{LOGLEVEL(y, s)} or\\
      \texttt{LOGLEVEL(y, p, q)} or\\
      \texttt{LOGLEVEL(y, p, d, q[, P, D, Q, s])} determines the log level specification with respect to various models. \texttt{y} is the observation data, \texttt{s} is the seasonal period. The output is \texttt{1} if no logs are needed, else \texttt{0}. The first form uses the airline model with specified seasonal period, second form uses ARMA models, and the third form uses SARIMA models.
\item \texttt{\textbf{oosforecast}}\\
      \texttt{[yf err SS] = OOSFORECAST(y, model, n1, h)} performs out-of-sample forecast. \texttt{y} is the data of dimension \texttt{p}${}\times{}$\texttt{n}, \texttt{model} is a \texttt{SSMODEL}, \texttt{n1} is the number of time points to exclude at the end, and \texttt{h} is the number of steps ahead to forecast, which can be an array. The output \texttt{yf} is the forecast obtained, \texttt{err} is the forecast error, and \texttt{SS} is the forecast error cumulative sum of squares.
\end{itemize}

\subsection{Stock element functions}

\begin{itemize}
\item \texttt{\textbf{fun\_arma}}\\
      \texttt{[fun grad psi] = FUN\_ARMA(p, q)} creates matrix update functions for ARMA models.
\item \texttt{\textbf{fun\_cycle}}\\
      \texttt{[fun grad psi] = FUN\_CYCLE()} creates matrix update functions for cycle components.
\item \texttt{\textbf{fun\_dupvar}}\\
      \texttt{[fun grad psi] = FUN\_DUPVAR(p, cov, d[, name])} creates matrix update functions for duplicated variances.
\item \texttt{\textbf{fun\_genair}}\\
      \texttt{[fun grad psi] = FUN\_GENAIR(nparam, nfreq, freq)} creates matrix update functions for generalized airline models.
\item \texttt{\textbf{fun\_homovar}}\\
      \texttt{[fun grad psi] = FUN\_HOMOVAR(p, cov, q[, name])} creates matrix update functions for homogeneous variances.
\item \texttt{\textbf{fun\_interlvar}}\\
      \texttt{[fun grad psi] = FUN\_INTERLVAR(p, q, cov[, name])} creates matrix update functions for q-interleaved variances.
\item \texttt{\textbf{fun\_mvcycle}}\\
      \texttt{[fun grad psi] = FUN\_MVCYCLE(p)} creates matrix update functions for multivariate cycle component.
\item \texttt{\textbf{fun\_oneoverf}}\\
      \texttt{[fun grad psi] = FUN\_ONEOVERF(m)} creates matrix update functions for 1/f noise.
\item \texttt{\textbf{fun\_sarimahtd}}\\
      \texttt{[fun grad psi] = FUN\_SARIMAHTD(p, d, q, P, D, Q, s, gauss)} creates matrix update functions for SARIMA with Hillmer-Tiao decomposition and non-Gaussian irregular.
\item \texttt{\textbf{fun\_sarma}}\\
      \texttt{[fun grad psi] = FUN\_SARMA(p, q, P, Q, s)} creates matrix update functions for seasonal ARMA models.
\item \texttt{\textbf{fun\_spline}}\\
      \texttt{[fun grad psi] = FUN\_SPLINE(delta)} creates matrix update functions for the cubic spline models.
\item \texttt{\textbf{fun\_var}}\\
      \texttt{[fun grad psi] = FUN\_VAR(p, cov[, name])} creates matrix update functions for variances.
\item \texttt{\textbf{fun\_wvar}}\\
      \texttt{[fun grad psi] = FUN\_WVAR(p, s[, name])} creates matrix update functions for W structure variances.
\item \texttt{\textbf{mat\_arima}}\\
      \texttt{[Z T R P1 Tmmask Rmmask P1mmask] = MAT\_ARIMA(p, d, q, mean)} creates base matrices for ARIMA models.
\item \texttt{\textbf{mat\_arma}}\\
      \texttt{[Z T R P1 Tmmask Rmmask P1mmask] = MAT\_ARMA(p, q, mean)} or\\
      \texttt{[T R P1 Tmmask Rmmask P1mmask] = MAT\_ARMA(p, q, mean)} creates base matrices for ARMA models.
\item \texttt{\textbf{mat\_commonlvls}}\\
      \texttt{[Z T R a1 P1] = MAT\_COMMONLVLS(p, A, a)} creates base matrices for common levels models.
\item \texttt{\textbf{mat\_dummy}}\\
      \texttt{[Z T R] = MAT\_DUMMY(s[, fixed])} creates base matrices for dummy seasonal component.
\item \texttt{\textbf{mat\_dupvar}}\\
      \texttt{[m mmask] = MAT\_DUPVAR(p, cov, d)} creates base matrices for duplicated variances.
\item \texttt{\textbf{mat\_homovar}}\\
      \texttt{[H Q Hmmask Qmmask] = MAT\_HOMOVAR(p, cov, q)} creates base matrices for homogeneous variances.
\item \texttt{\textbf{mat\_hs}}\\
      \texttt{[Z T R] = MAT\_HS(s)} creates base matrices for Harrison and Stevens seasonal component.
\item \texttt{\textbf{mat\_interlvar}}\\
      \texttt{[m mmask] = MAT\_INTERLVAR(p, q, cov)} creates base matrices for q-interleaved variances.
\item \texttt{\textbf{mat\_lpt}}\\
      \texttt{[Z T R] = MAT\_LPT(d, stochastic)} creates base matrices for local polynomial trend or integrated random walk models.
\item \texttt{\textbf{mat\_mvcycle}}\\
      \texttt{[Z T Tmmask R] = MAT\_MVCYCLE(p)} creates base matrices for multivariate cycle components.
\item \texttt{\textbf{mat\_mvdummy}}\\
      \texttt{[Z T R] = MAT\_MVDUMMY(p, s[, fixed])} creates base matrices for multivariate dummy seasonal components.
\item \texttt{\textbf{mat\_mvhs}}\\
      \texttt{[Z T R] = MAT\_MVHS(p, s)} creates base matrices for multivariate Harrison and Stevens seasonal components.
\item \texttt{\textbf{mat\_mvintv}}\\
      \texttt{[Z Zdmmask Zdvec T R] = MAT\_MVINTV(p, n, type, tau)} creates base matrices for multivariate intervention components.
\item \texttt{\textbf{mat\_mvllt}}\\
      \texttt{[Z T R] = MAT\_MVLLT(p)} creates base matrices for multivariate local level trend models.
\item \texttt{\textbf{mat\_mvreg}}\\
      \texttt{[Z Zdmmask Zdvec T R] = MAT\_MVREG(p, x, dep)} creates base matrices for multivariate regression components.
\item \texttt{\textbf{mat\_mvtrig}}\\
      \texttt{[Z T R] = MAT\_MVTRIG(p, s[, fixed])} creates base matrices for multivariate trigonometric seasonal components.
\item \texttt{\textbf{mat\_reg}}\\
      \texttt{[Z Zdmmask Zdvec T R] = MAT\_REG(x[, dyn])} creates base matrices for regression components.
\item \texttt{\textbf{mat\_sarima}}\\
      \texttt{[Z T R P1 Tmmask Rmmask P1mmask] = MAT\_SARIMA(p, d, q, P, D, Q, s, mean)} or\\
      \texttt{[Z T R P1 Rmmask P1mmask] = MAT\_SARIMA(p, d, q, P, D, Q, s, mean)} creates base matrices for SARIMA models.
\item \texttt{\textbf{mat\_sarimahtd}}\\
      \texttt{[Z T R Qmat P1 Tmmask Rmmask Qmmask P1mmask] = MAT\_SARIMAHTD(p, d, q, P, D, Q, s)} creates base matrices for SARIMA with Hillmer-Tiao decomposition.
\item \texttt{\textbf{mat\_spline}}\\
      \texttt{[Z T Tdmmask Tdvec R] = MAT\_(delta)} creates base matrices for the cubic spline model.
\item \texttt{\textbf{mat\_sumarma}}\\
      \texttt{[Z T R P1 Tmmask Rmmask P1mmask] = MAT\_SUMARMA(p, q, D, s, mean)} creates base matrices for sum integrated ARMA models.
\item \texttt{\textbf{mat\_trig}}\\
      \texttt{[Z T R] = MAT\_TRIG(s[, fixed])} creates base matrices for trigonometric seasonal components.
\item \texttt{\textbf{mat\_var}}\\
      \texttt{[m mmask] = MAT\_VAR(p, cov)} creates base matrices for variances.
\item \texttt{\textbf{mat\_wvar}}\\
      \texttt{[m mmask] = MAT\_WVAR(p, s)} creates base matrices for W structure variances.
\item \texttt{\textbf{ngdist\_binomial}}\\
      \texttt{[matf logpf] = NGDIST\_BINOMIAL(k)} creates base functions for binomial distributions.
\item \texttt{\textbf{ngdist\_expfamily}}\\
      \texttt{[matf logpf] = NGDIST\_EXPFAMILY(b, d2b, id2bdb, c)} creates base functions for exponential family distributions.
\item \texttt{\textbf{ngdist\_multinomial}}\\
      \texttt{[matf logpf] = NGDIST\_MULTINOMIAL(h, k)} creates base functions for multinomial distributions.
\item \texttt{\textbf{ngdist\_negbinomial}}\\
      \texttt{[matf logpf] = NGDIST\_NEGBINOMIAL(k)} creates base functions for negative binomial distributions.
\item \texttt{\textbf{ngfun\_t}}\\
      \texttt{[fun psi] = NGFUN\_T([nu])} creates update functions for the t-distribution.
\item \texttt{\textbf{psi2err}}\\
      \texttt{distf = PSI2ERR(X)} is the update function for the general error distribution.
\item \texttt{\textbf{psi2mix}}\\
      \texttt{distf = PSI2MIX(X)} is the update function for the Gaussian mixture distribution.
\item \texttt{\textbf{psi2zmsv}}\\
      \texttt{distf = PSI2ZMSV(X)} is the update function for the zero-mean stochastic volatility model disturbance distribution.
\item \texttt{\textbf{x\_ee}}\\
      \texttt{X\_EE(Y, M, d)} generates Easter effect variable.
\item \texttt{\textbf{x\_intv}}\\
      \texttt{X\_INTV(n, type, tau)} generates intervention variables.
\item \texttt{\textbf{x\_ly}}\\
      \texttt{X\_LY(Y, M)} generates leap-year variable.
\item \texttt{\textbf{x\_td}}\\
      \texttt{X\_TD(Y, M, td6)} generates trading day variables.
\end{itemize}

\subsection{Miscellaneous helper functions}
\label{app:function:helper}

\begin{itemize}
\item \texttt{\textbf{diaginprod}}\\
      \texttt{y = DIAGINPROD(A, x1[, x2])} calculates the inner product of a vector sequence. \texttt{A} is a $m\times m$ square matrix. \texttt{x1} and \texttt{x2} are $m\times n$ matrices representing a set of $n$ column vectors each of size $m\times1$. The $(i,j)$th element of \texttt{y} is equal to \texttt{(x1(:, j)${}-{}$x2(:, i))$^T$A(x1(:, j)${}-{}$x2(:, i))}.
\item \texttt{\textbf{f\_alpha2arma}}\\
      \texttt{[xAR xMA] = F\_ALPHA2ARMA(n, m, alpha, sigma2)} generates 1/f noise by AR and MA approximations. \texttt{n} is length of series to generate, \texttt{m} is number of terms to use in the approximation, \texttt{alpha} is the exponent of inverse frequency of the 1/f noise, and \texttt{sigma2} the variance of the 1/f noise. The output is the 1/f noise generated by AR and MA approximation respectively.
\item \texttt{\textbf{meancov}}\\
      \texttt{[m C] = MEANCOV(x)} calculates the mean and covariance of each vector set in a sequence. \texttt{x} is a $m\times n\times N$ matrix representing $n$ vector sets each of size $N$ containing $m\times1$ vectors. \texttt{m} is a $m\times n$ matrix, the mean of each of the $n$ vector sets. \texttt{C} is a $m\times m\times n$ matrix, the covariance of each of the $n$ vector sets.
\item \texttt{\textbf{randarma}}\\
      \texttt{RANDARMA(n, r)} generates random ARMA polynomials. \texttt{n} is the polynomial degree and \texttt{r} is root magnitude range.
\item \texttt{\textbf{setopt}}\\
      \texttt{SETOPT(optname1, optvalue1, optname2, optvalue2, \ldots)} sets the global analysis settings and returns a structure containing the new settings, thus a call with no arguments just returns the current settings. The available options are: \texttt{'disp'} can be set to \texttt{'off'}, \texttt{'notify'}, \texttt{'final'} or \texttt{'iter'}. \texttt{tol} is the tolerance. \texttt{fmin} is the minimization algorithm to use (\texttt{'bfgs'} or \texttt{'simplex'}). \texttt{maxiter} is the maximum number of iterations. \texttt{nsamp} is the number of samples to use in simulation. Note that the same form of argument sequence can be used at the end of most data analysis functions to specify one time overrides.
\item \texttt{\textbf{ymarray}}\\
      \texttt{[Y M] = YMARRAY(y, m, N)} generates an array of years and months from specified starting year \texttt{y}, starting month \texttt{m}, and number of months \texttt{N}.
\end{itemize}

\end{document}